\documentclass{tcibook}
\usepackage{fancyhea}
\usepackage{work}
\usepackage{bm}       
\usepackage{graphicx}
\usepackage{hyperref,amssymb,color}      

\usepackage{chngcntr}
\counterwithout{figure}{chapter}

\newcommand{\nc}{\newcommand}  

\def\beq{\begin{equation}}
\def\eeq#1{\label{#1}\end{equation}}
\def\eeqn{\end{equation}}

\newenvironment{Eqnarray}%
   {\arraycolsep 0.14em\begin{eqnarray}}{\end{eqnarray}}
\def\beqa{\begin{Eqnarray}}
\def\eeqa#1{\label{#1}\end{Eqnarray}}
\def\eeqan{\end{Eqnarray}}

\nc{\ra}{\rightarrow}  
\nc{\slsh}{\slash\hspace*{-0.22cm}}
\def\Re{{\cal R \mskip-4mu \lower.1ex \hbox{\it e}\,}}
\def\Im{{\cal I \mskip-5mu \lower.1ex \hbox{\it m}\,}}

\nc{\vev}[1]{ \left\langle {#1} \right\rangle }
\nc{\bra}[1]{ \langle {#1} | }
\nc{\ket}[1]{ | {#1} \rangle }
\nc{\fb}{\,{\rm fb}^{-1}}
\nc{\ev}{{\rm eV}}
\nc{\kev}{{\rm keV}}
\nc{\Mev}{{\rm MeV}}
\nc{\gev}{{\rm GeV}}
\nc{\tev}{{\rm TeV}}
\nc{\mev}{{\rm MeV}}

\def\del{\partial}
\def\Dslash{\not{\hbox{\kern-4pt $D$}}}
\def\dslash{\not{\hbox{\kern-2pt $\del$}}}
\def\pslash{\not{\hbox{\kern-2pt $p$}}}
\def\ETmiss{ \not{\hbox{\kern-4pt $E$}}_T }

\def\msb{{\bar{\ssstyle M \kern -1pt S}}}

\begin{document}

\def\bibname{References}

\bibliographystyle{utphys}  

\raggedbottom

\pagenumbering{roman}

\parindent=0pt
\parskip=8pt
\setlength{\evensidemargin}{0pt}
\setlength{\oddsidemargin}{0pt}
\setlength{\marginparsep}{0.0in}
\setlength{\marginparwidth}{0.0in}
\marginparpush=0pt


\pagenumbering{arabic}

\renewcommand{\chapname}{chap:cc}
\renewcommand{\chapterdir}{.}
\renewcommand{\arraystretch}{1.25}
\addtolength{\arraycolsep}{-3pt}


\newcommand{\red}[1]{\textcolor{red}{#1}}
\newcommand{\fs}[1]{\textcolor{red}{[\textbf{FS:}~#1]}}

\newcommand*\mystrut[1]{\vrule width0pt height0pt depth#1\relax}                                

\def\hmpcinv{\,h\,{\rm Mpc^{-1}}}
\def\hinvmpc{\,h^{-1}{\rm Mpc}}

\newcommand{\sigabs}{\sigma_{11,\mathrm{abs}}}
\newcommand{\Planck}{\emph{Planck}}
\newcommand{\Euclid}{\emph{Euclid}}
\newcommand{\WFIRST}{\emph{WFIRST}}
\newcommand\apj{{ApJ}}
\newcommand\aj{{AJ}}
\newcommand\apjs{{ApJ S.}}
\newcommand\mnras{{MNRAS}}
\newcommand\aap{ {A \& A}}
\newcommand{\prd}{Phys. Rev. D}

\newcommand{\dragan}[1]{\textcolor{red}{[{\bf DH}: #1]}}
\newcommand{\kyle}[1]{\textcolor{blue}{[{\bf Kyle}: #1]}}

\chapter*{Growth of Cosmic Structure:\\[0.1cm] Probing Dark Energy Beyond Expansion} 
\renewcommand*\thesection{\arabic{section}}
\label{chap:mag}

\begin{center}\begin{boldmath}



\begin{center}

\begin{large} {\bf Conveners: Dragan Huterer, David Kirkby} \end{large}

Rachel Bean,
Andrew Connolly,
Kyle Dawson,
Scott Dodelson,
August Evrard,
Bhuvnesh Jain,
Michael Jarvis,
Eric Linder, 
Rachel Mandelbaum, 
Morgan May,
Alvise Raccanelli,
Beth Reid,
Eduardo Rozo,
Fabian Schmidt,
Neelima Sehgal,
An\v{z}e Slosar,
Alex van Engelen,
Hao-Yi Wu,
Gongbo Zhao

\end{center}



\end{boldmath}\end{center}

\section{Executive Summary}

The quantity and quality of cosmic structure observations have greatly
accelerated in recent years, and further leaps forward will be
facilitated by imminent projects.  These will enable us to map the
evolution of dark and baryonic matter density fluctuations over cosmic
history.  The way that these fluctuations vary over space and time is
sensitive to several pieces of fundamental physics: the primordial
perturbations generated by GUT-scale physics; neutrino masses and
interactions; the nature of dark matter and dark energy.  We focus on
the last of these here: the ways that combining probes of growth with
those of the cosmic expansion such as distance-redshift relations will
pin down the mechanism driving the acceleration of the Universe.

One way to explain the acceleration of the Universe is invoke dark
energy parameterized by an equation of state $w$. Distance
measurements provide one set of constraints on $w$, but dark energy
also affects how rapidly structure grows; the greater the
acceleration, the more suppressed the growth of structure.  Upcoming
surveys are therefore designed to probe $w$ with direct observations
of the distance scale and the growth of structure, each complementing
the other on systematic errors and constraints on dark energy.  A
consistent set of results will greatly increase the reliability of the
final answer.

Another possibility is that there is no dark energy, but that General
Relativity does not describe the laws of physics accurately on large
scales. While the properties of gravity have been measured with
exquisite precision at stellar system scales and densities, within our
solar system and by binary pulsar systems, its properties in different
environments are poorly constrained. To fully understand if General
Relativity is the complete theory of gravity we must test gravity across a
spectrum of scales and densities.  Rapid developments in gravitational
wave astronomy and numerical relativity are directed at testing
gravity in the high curvature, high density regime. Cosmological
evolution provides a polar opposite test bed, probing how gravity
behaves in the lowest curvature, low density environments.

There are a number of different implementations of astrophysically relevant
modifications of gravity. Generically, the models are able to reproduce the
distance measurements but at the cost of altering the growth of structure. In
particular, as detailed below, the Poisson equation relating over-densities to
gravitational potentials is altered, and the potential that determines the
geodesics of relativistic particles (such as photons) differs from the
potential that determines the motion of non-relativistic particles.  Upcoming
surveys will exploit these differences to determine whether the acceleration
of the Universe is due to dark energy or to modified gravity.

To realize this potential, both wide field imaging and spectroscopic
redshift surveys play crucial roles.  Projects including DES, eBOSS,
DESI, PFS, LSST, \Euclid, and \WFIRST\ are in line to map more than a 1000
cubic-billion-light-year volume of the Universe. These will map the
cosmic structure growth rate to 1\% in the redshift range $0<z<2$,
over the last 3/4 of the age of the Universe.

\section{Introduction: Why Measuring Growth is Interesting}

The standard cosmological model posits that the largest structures
that we observe today --- galaxies and clusters of galaxies --- grew
out of small initial fluctuations that were seeded during the phase of
inflationary expansion, some $10^{-35}$ seconds after the Big
Bang. Subsequently these fluctuations grew under the influence of
gravity. Most of the growth occurred after the decoupling of photons
and electrons, some 350,000 years after the Big Bang, when a sudden
drop in Jeans mass, as well as the fact that the Universe was mostly
matter-dominated at that point, allowed the galactic-size structures
to grow unimpeded.

In the currently favored cosmological model, where most of the matter
budget is dominated by the slow-moving massive particles (the ``cold
dark matter'' or CDM), the smaller structures form first, while the
largest structures form the latest. Therefore, objects that are of the
most interest to cosmologists, galaxies and clusters of galaxies, form
at recent times and, in some cases, are still forming today. Hence,
observations in various wavelengths can probe the full evolution of
the formation of structure in the Universe, from when the first
objects formed until today.

Observations of the growth of structure provide a wealth of
information about dark matter and dark energy. In particular, the
scaling of the amplitude of growth vs.\ cosmic time --- the so-called
growth function --- sensitively constrains dark energy parameters in a
way that is complementary to distance measurements. The temporal
evolution of the growth is now readily observed by measuring the
clustering of galaxies at multiple redshifts, and in the near future
gravitational lensing has the potential to measure the same quantity
but with the added advantage that it is directly sensitive to the
growth of dark matter structures (as opposed to galaxies or other
baryonic tracers such as hydrogen in the inter-galactic
medium). Additionally, the number counts of clusters of galaxies, as a
function of their mass and redshift, provide another excellent probe
of cosmological parameters. Our ability to observe and model both the
growth and the cluster counts have significantly matured over the past
decade, and these two probes now provide constraints on dark energy
that are complementary to distance measurements by type Ia supernovae,
baryon acoustic oscillations (BAO; which encode geometrical aspects of
the clustering of galaxies), and the cosmic microwave background
(CMB).

Over the next 10--20 years, we expect a wealth of new observations that
include ground imaging surveys (e.g.\ DES and LSST), redshift surveys
(e.g.\ eBOSS, PFS and DESI) and space surveys (e.g.\ \Euclid\ and
\WFIRST). The combination of these observations will provide high-precision
measurements of the growth of structure out to redshift of a few and across
most of the sky. These measurements will, in turn, strongly constrain the
equation of state of dark energy and, more generally, the expansion history of
the Universe (discussed in the Snowmass-2013 paper on Distances
\cite{Kim:2013nma}) over the past $\sim$10 billion years.

The growth of structure is particularly sensitive as a probe of
modified-gravity explanations for the accelerating Universe, and has already
been used to impose constraints on the extensions of, and modifications to,
General Relativity (GR). The effectiveness of the growth of structure in this
regard is maximized when it is combined with distance measurements. Roughly
speaking, accurate measurements of the distances predict the growth of
structure when GR is assumed to be correct and unmodified. Therefore,
independent precision measurements of the growth of structure test whether GR
adequately describes the late-time expansion of the Universe. Such tests are
paramount to our understanding of dark energy and may lead to fundamental
discoveries of physics at large scales, and this makes the growth of structure
a very important probe of the Universe.

Good complementarity, redundancy, and control of the systematic errors are
keys in making the growth of structure observations reach their full
potential. Photometric and spectroscopic surveys are particularly
complementary in various aspects of their observational strategies; moreover,
spectroscopic surveys play an additional key role of calibrating the
photometric redshifts obtained from galaxy colors. Multiple observations of
the same sky coverage may be useful for this reason, while non-overlapping
observations help reduce cosmic variance. Finally, numerical (N-body)
simulations have an extremely important role of providing theoretical
predictions for the growth of structure in the quasi-linear regime (roughly
10--50 megaparsecs) and especially in the non-linear regime (scales less than
about 10 megaparsecs).

The paper is organized in follows. In Sec.~\ref{sec:prelim} we define
what precisely we mean by the growth of structure, and broadly
illustrate constraints on it from future surveys.  In
Sec.~\ref{sec:DE-MG} we discuss how the growth of structure probes the
dark-energy and modified-gravity explanations for the acceleration of
the Universe.  In Sec.~\ref{sec:probes} we discuss in some detail
several of the most promising probes of the growth of structure --
clustering of galaxies in spectroscopic surveys, counts of galaxy
clusters, and weak gravitational lensing.  Finally in
Sec.~\ref{sec:sim} we discuss the very important role of simulations
in theoretically predicting growth on non-linear scales.

\section{Preliminaries and Definitions}\label{sec:prelim}

\begin{figure*}[t]
\begin{center}
\includegraphics[width=1\textwidth]{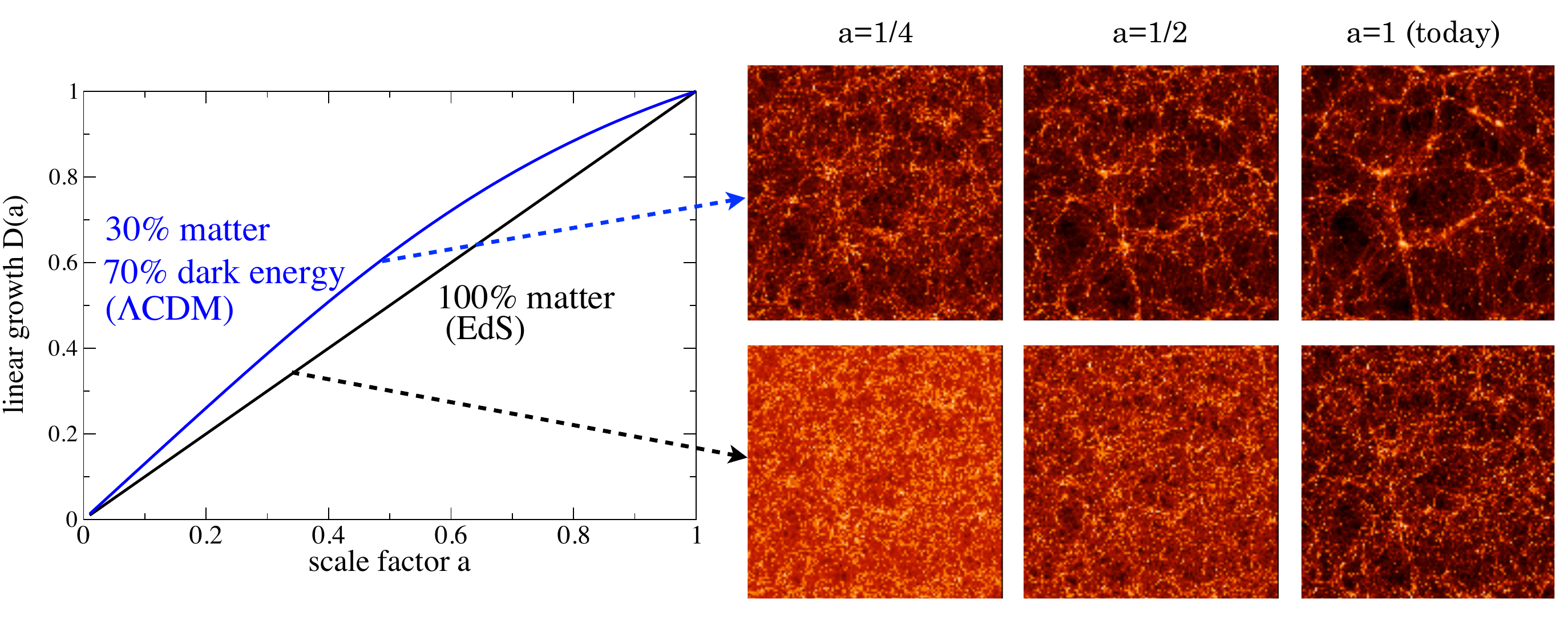}
\end{center}
\caption{Growth of structure at large spatial scales in the universe. {\bf Left
  panel:} Because dark energy suppresses the growth of structure, the linear
  growth $D(a)$, which is normalized to unity today, had to be larger in the
  past in the currently favored model with dark energy ($\Lambda$CDM; blue
  line) than in the Einstein-de Sitter model (EdS; black line) which has
  matter only and no dark energy. {\bf Right panel:} snapshots from numerical
  (N-body) simulations by the Virgo consortium \cite{Jenkins:1997en}, showing
  larger amplitude of density fluctuations in the past in $\Lambda$CDM (top
  row) than in the EdS model (bottom row) given an approximately fixed
  amount of clustering today. Accurate measurements of the clustering as a
  function of spatial scale and cosmic time can therefore stringently
  constrain the cosmological model.  }
\label{fig:Da_basic}
\end{figure*}

In the linear theory --- valid at sufficiently early times and sufficiently
large spatial scales, when the fluctuations in the matter energy density
$\rho_M$ are much less than unity --- the matter density contrast
$\delta=\delta\rho_M/\rho_M$ evolves independently of the spatial scale
$k$. The growth of fluctuations in time (well within the Hubble radius) can be
obtained by solving the equation
\begin{equation}
\ddot\delta + 2H\dot\delta-4\pi G\rho_M \delta = 0,
\label{eq:growth}
\end{equation}
where $H$ is the Hubble parameter, and dots are derivatives with respect to
time $t$.  Therefore, in standard General Relativity and in the linear regime
($|\delta| \ll 1$), obtaining the linear growth of fluctuations as a function
of time is straightforward given the composition and the expansion rate of the
Universe.  More commonly, one works in terms of the scale factor $a$, where
$d\ln a = H dt$, and $a=0$ ($a=1$) at the Big Bang (today). One can then
define the {\it linear growth function} $D(a)$ via
\begin{equation}
\delta(a)=D(a)\delta(a=1),
\end{equation}
or equivalently in redshift $z$ where $1+z=1/a$. If General Relativity is
replaced by some modified gravity theory, then Eq.~(\ref{eq:growth}) changes:
the evolution of growth needs to be re-derived in the new theory, and the
linear growth rate $D$ may then depend on scale $k$ as well. The left panel of
Fig.~\ref{fig:Da_basic} illustrates the linear growth function for two
representative cosmological models, and the right panel provides snapshots
from numerical simulation further illustrating the suppressed growth of
structure in the presence of dark energy.

\begin{figure*}[t]
\begin{center}
\includegraphics[width=0.7\textwidth]{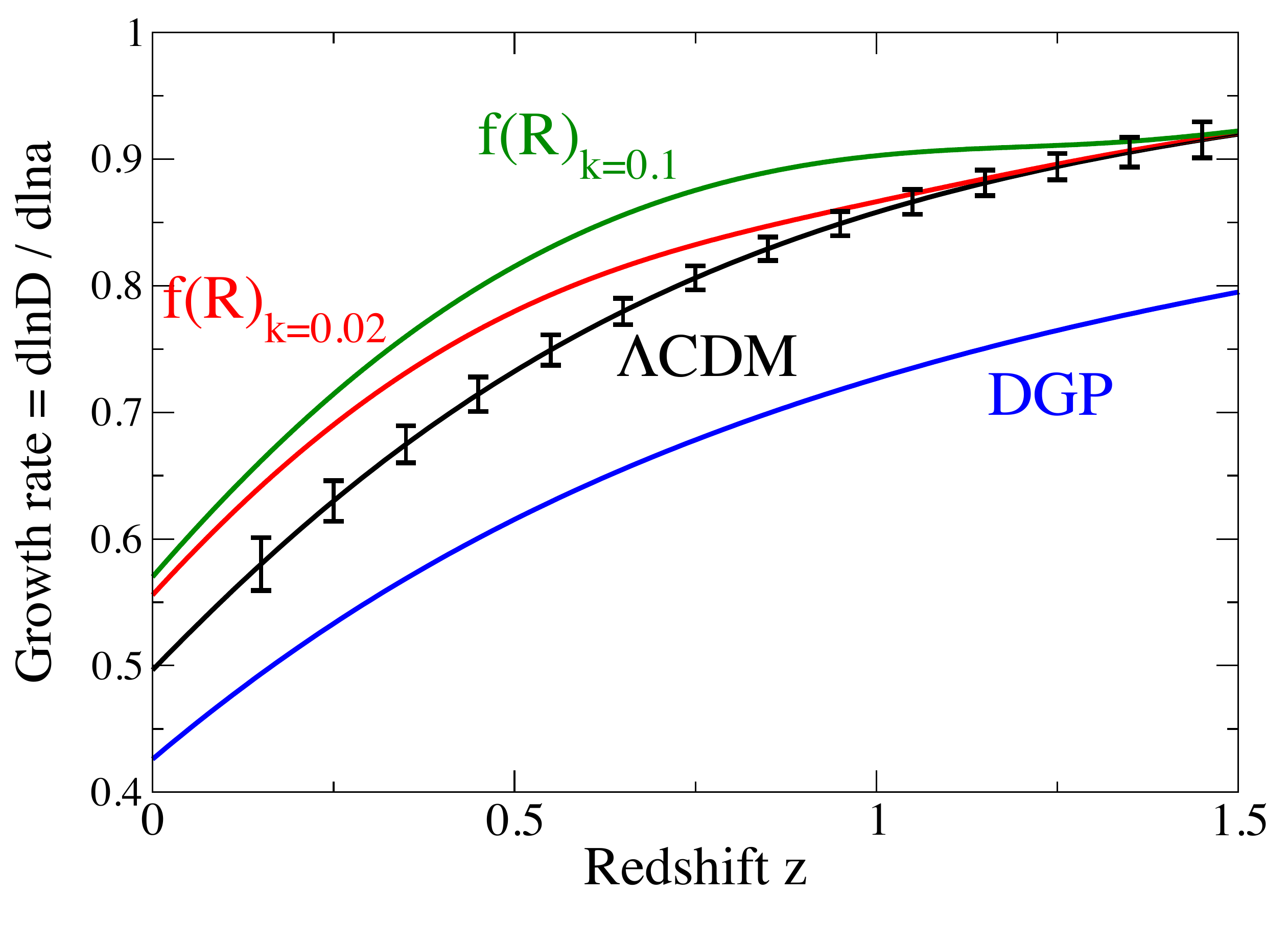} 
\end{center}
\vspace{-0.6truecm}
\caption{Constraints on the growth of density fluctuations in the Universe
  with errors projected from a future survey designed with DESI
  specifications. The curves show the derivative of the logarithmic growth
  with respect to the logarithmic scale factor --- a quantity readily measured
  from the clustering of galaxies in redshift space --- as a function of
  redshift. We show theory predictions for the $\Lambda$CDM model, as well as
  for two modified-gravity models: the Dvali-Gabadadze-Porrati braneworld
  model \cite{DGP} and the $f(R)$ modification to the Einstein action
  \cite{0905.2962}. Because growth in the $f(R)$  models is
  generically scale-dependent, we show predictions at two wavenumbers,
  $k=0.02\hmpcinv$ and $k=0.1\hmpcinv$. LSST projects to impose constraints of similar excellent quality on the growth function $D(a)$.
}
\label{fig:killer}
\end{figure*}

It is useful to make a connection to other quantities commonly used to probe
growth. First of all, we can define the matter {\it power spectrum} $P(k)$ as
the Fourier transform of the 2-point correlation function, $\langle
\delta_{\vec{k}}\,\delta_{\vec{k}'}^*\rangle = (2\pi)^3\,
\delta^{(3)}(\vec{k}-\vec{k}')\,P(k)$. Note that $P(\vec{k})=P(k)$ due to
isotropy of the Universe and the scale-factor- (or redshift-) dependence of
both $\delta$ and $P$ is implicit.  A commonly used quantity is the rms {\it
  amplitude of mass fluctuations}, whose square (i.e.\ the variance of mass
fluctuations $\sigma_R^2$) is given by the integral over the power spectrum
defined {\it in linear theory}
\begin{equation} 
\sigma^2_R(a) = \int_0^\infty \frac{k^3 P_{\rm linear}(k, a)}{2\pi^2}\,
W^2(kR)\, d\ln k 
\end{equation}
where $W(x)=3j_1(x)/x$ is the Fourier transform of the real-space window function.
The quantity $\sigma_R(a)$ encodes the amount of matter fluctuations averaged over a sphere
of radius $R$ at redshift $z$, assuming that the fluctuations are fully
linear (thus Eq.~(\ref{eq:growth}) is valid). A common choice to describe the normalization of
the fluctuations in the Universe {\it today} is $\sigma_8\equiv
\sigma_{8\hinvmpc}(a=1)$. Measurements of the redshift-dependence of
$\sigma_8$ are sometimes quoted as probes of the growth function $D(a)$, since
$\sigma_8(a)=\sigma_8D(a)$.

Figure \ref{fig:killer} shows an example of constraints from the growth of
structure, shown for only one of the cosmological probes --- the
redshift-space distortions (RSD), which will be further discussed in
Sec.~\ref{sec:rsd}. This probe is sensitive to the derivative of the logarithm
of the growth function with respect to logarithm of
the cosmic scale\footnote{More precisely, the RSD are sensitive to $\sigma_8(a)$
  times this quantity, but we ignore this subtle distinction for the moment.};
we thus show the quantity
\begin{equation}
f(a)\equiv \frac{d\ln D}{d\ln a}
\label{eq:f}
\end{equation}
vs.\ the redshift $z\equiv 1/a-1$.  We show theory predictions for the
currently favored cosmological constant plus cold dark matter ($\Lambda$CDM)
model, as well as for two modified-gravity models, the Dvali-Gabadadze-Porrati
braneworld model (DGP; \cite{DGP}), and the $f(R)$ modification to Einstein
action from Ref.~\cite{0905.2962} with $c=3$. Because growth in the $f(R)$
models is generically scale-dependent, we show predictions at two
wavenumbers, $k=0.02\hmpcinv$ and $k=0.1\hmpcinv$.  The $f(R)$ model --- which
is usually challenging to distinguish from GR because it can have the
expansion history mimicking the $\Lambda$CDM model ($w$ is within 1\% of $-1$)
and can have a growth function identical to $\Lambda$'s at high redshift ---
can clearly be distinguished from $\Lambda$CDM using growth data from future
surveys such as eBOSS, DESI, \Euclid, or \WFIRST.  The DGP model can be
distinguished even more readily by measuring both the expansion history as
well as growth of structure in the Universe.

\section{Dark Energy and Modified Gravity}
\label{sec:DE-MG}

Over the past decade, the $\Lambda$CDM paradigm has passed all observational
tests, firmly establishing it as our cosmological ``standard model''.
However, it is clearly of crucial importance to test this paradigm, given that
it involves two unknown ingredients (dark matter and $\Lambda$), and given the
lack of theoretical motivation for the value of the putative cosmological
constant. Growth of structure offers a broad range of probes of dark energy
which in principle cover three orders of magnitude in length scale, and one
order of magnitude in time or scale factor.  In order to convincingly rule out
alternatives to the cosmological constant, we need to cover this range of
scales and redshifts.  Large-scale structure also provides model-independent
tests of gravity on Mpc scales and above, extending Solar System tests by ten
orders of magnitude in length scale.

In this section we briefly discuss the theoretical underpinnings of the tests
of the accelerating Universe with the growth of cosmic structure. Modified
gravity in particular is covered in much more detail in the Snowmass-2013
paper on Novel Probes of Gravity and Dark Energy \cite{Jain:2013uma}.

\subsection{Dark Energy}

The physics behind the observed accelerated expansion of the Universe
is widely recognized as one of the most profound outstanding problems in
fundamental science.  When interpreted in terms of our current understanding
of gravity, Einstein's General Relativity, this requires
adding an additional, exotic component to the cosmic energy budget with
a negative pressure, which we now refer to as ``Dark Energy''.  

The minimalist explanation is to invoke a very small cosmological constant.
However, within our current understanding of quantum theory, such a value of
the cosmological constant is extremely unnatural.  Instead, one can invoke a very light scalar field,
whose potential energy then drives the accelerated expansion of the Universe
(this of course does not solve the cosmological constant problem).  In fact,
this is precisely the mechanism surmised to having produced inflation, an
epoch of extremely rapid expansion in the very early Universe which allowed
the Universe to grow to its observed large size, and which provided the seed
fluctuations for the structure within the Universe.  While we have no good
theoretical framework to connect the very early and late time epochs of
acceleration, inflation can be seen as a tantalizing hint that the late time
acceleration might be transitory and thus not be due to a cosmological
constant.  Observationally, we can distinguish this case by measuring the
equation of state parameter $w\equiv p_{\rm DE}/\rho_{\rm DE}$, which is exactly $-1$ for a cosmological
constant but slightly larger for a ``slowly rolling'' scalar field.  The
parameter $w$ affects both the expansion history (geometry) of the Universe as
well as the growth of structure within it.

A smooth dark energy component is completely described by its equation of
state as function of time, $w(t)$.  For this reason, {\it there is a
  consistency relation between the expansion history (for example, as measured by
  type~Ia supernovae or BAO) and the growth of structure as measured by weak
  lensing, galaxy clusters, and redshift space distortions}.
In simplest terms, this can be illustrated as the
consistency between the linear perturbations $\delta(t)$ and the expansion
history determined via the Hubble parameter $H(t)$ --- in standard, unmodified
GR, they need to satisfy the linear growth equation (Eq.\ \ref{eq:growth}). Thus,
measurements of the large-scale structure can unambiguously falsify the smooth
dark energy paradigm.

Going beyond the simplest models of dark energy, one can consider fluctuations
in the dark energy density, which require (at least at some point in time)
a value of $w$ significantly different from $-1$.  The amplitude of these
fluctuations is controlled by the sound speed of the scalar degree of freedom.  
Further, one can allow for a coupling of the scalar field to other 
components of matter.  While a coupling to ordinary matter and radiation
is constrained to be small (but see below), a coupling to dark matter
is much less constrained and can only be probed through large-scale
structure.  These possibilities go beyond the simplest models of dark energy;
however, they are still allowed by the data and, if detected, would allow for
rich insights into the physics of dark energy and dark matter.

\subsection{Modified Gravity}

As a fundamental alternative to dark energy, one can ask whether the
acceleration of the Universe is caused by a modification of gravity on large
scales, i.e. departure from GR, rather than an exotic form of
energy.  This possibility has generated a significant amount of theoretical
work over the past decade; it furthermore provides strong motivation to search
for and constrain modifications to GR using cosmological observations.
However, modifying GR on large scales in a consistent way is extremely
difficult, due to both theoretical issues and a broad set of observational
constraints.  In particular, any theory of gravity has to reduce to GR within
the Solar System to satisfy stringent local tests of gravity.  Further, the
cosmic microwave background and the Big Bang nucleosynthesis provide constraints
in the early Universe.  Both of these constraints can be satisfied by invoking
non-linear ``screening mechanisms'' which restore GR in high density regions.

Several mechanisms that achieve this have been proposed in the literature; they
 manage to hide a light scalar degree of freedom with
gravitational-strength coupling to matter in high density regions.  They
operate by either making the field massive in high-density regions (chameleon
mechanism), or suppressing its coupling through non-linear interactions
(Vainshtein and symmetron mechanisms).  When placing constraints on gravity
models using structure in the non-linear regime it is then important to take
into account the effects of these screening mechanisms.  For example, in
models with chameleon screening, the abundance of massive halos is not
strongly enhanced over $\Lambda$CDM as predictions from linear growth of
structure would suggest.  On the other hand, the screening mechanisms can lead
to unique signatures of their own in large-scale structure.  These screening
mechanisms are also relevant for models with a dark energy coupled to ordinary
matter.\\

\subsection{Distinguishing Between Dark Energy and Modified Gravity}
 
In order to distinguish between a smooth dark energy component and less
minimal models such as coupled dark energy or modified gravity, it is crucial
to measure the growth of structure in addition to the expansion history.  This
is because any given expansion history predicted by a modified gravity
model could be emulated by a smooth dark energy component.

Observations of the large-scale structure can roughly be divided into two
regimes:
\begin{itemize}
\item On {\it large scales}, fractional density perturbations are much less
  than one and are amenable to a perturbative treatment, so that the
  theoretical predictions for the growth are very accurate.  In the context of
  dark energy and modified gravity, this regime is useful since one can
  parametrize the stress-energy content of the Universe as well as the
  relation between stress energy and metric potentials, that is, gravity.
  Thus, in this regime one can place model-independent constraints on general
  dark energy models and modifications to gravity.

\item On {\it smaller scales} ($\lesssim 10$~Mpc), density fluctuations become
  non-linear, and the perturbative treatment breaks down.  Nevertheless, on
  scales larger than a few Mpc gravity is still the only relevant force, and
  quantitative predictions can be made through N-body simulations which are
  discussed in Sec.~\ref{sec:sim}.  Thus, this
  regime can still be used to probe gravity and dark energy.  While clearly
  much more challenging to model and confront with data, the bulk of the
  information delivered by growth is in this regime, so it is
  essential to make use of it.  In case of modifications to gravity, it is
  important to take into account the screening mechanisms as well.
\end{itemize}

\begin{figure}[t]
\begin{center}
\includegraphics[width=0.49\textwidth]{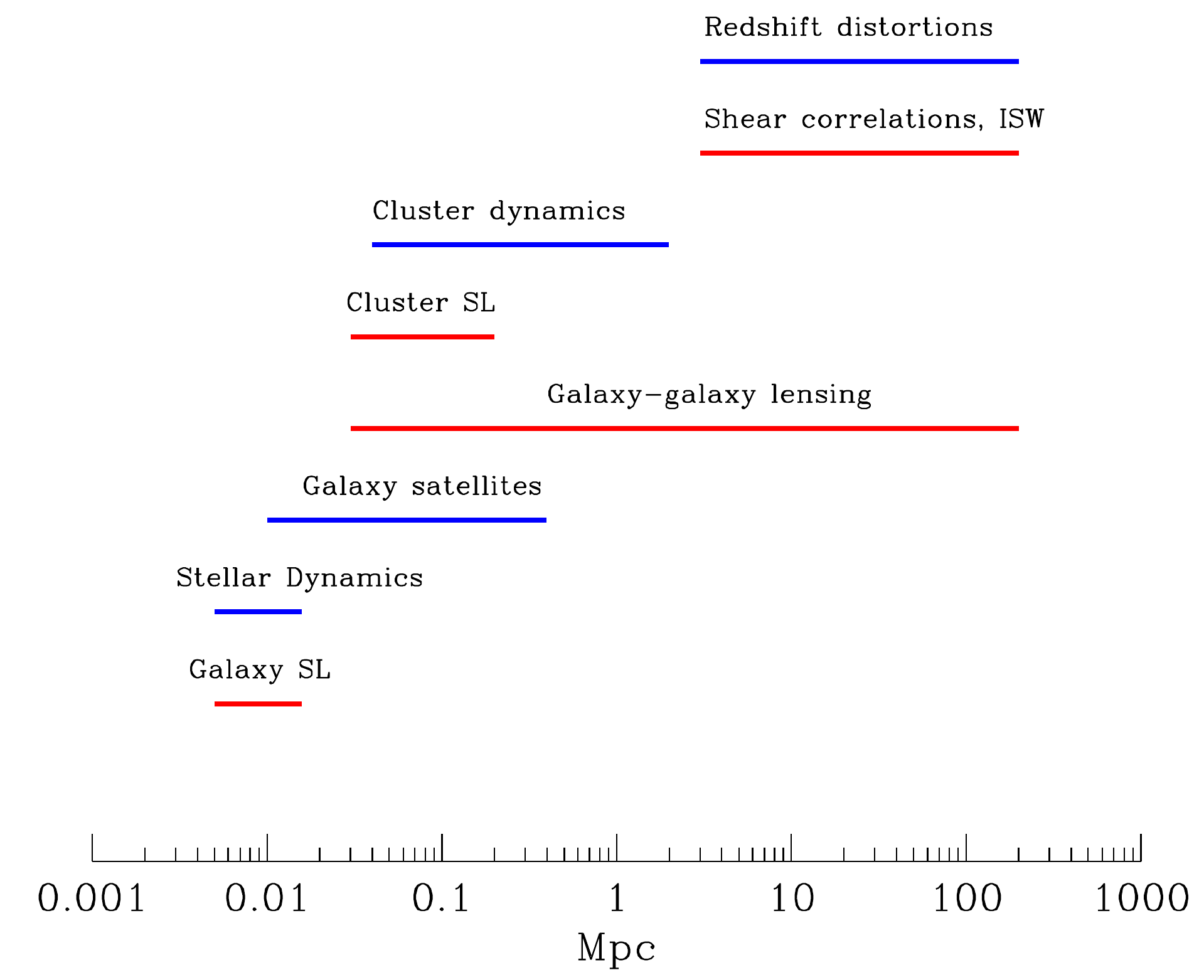}
\includegraphics[width=0.50\textwidth]{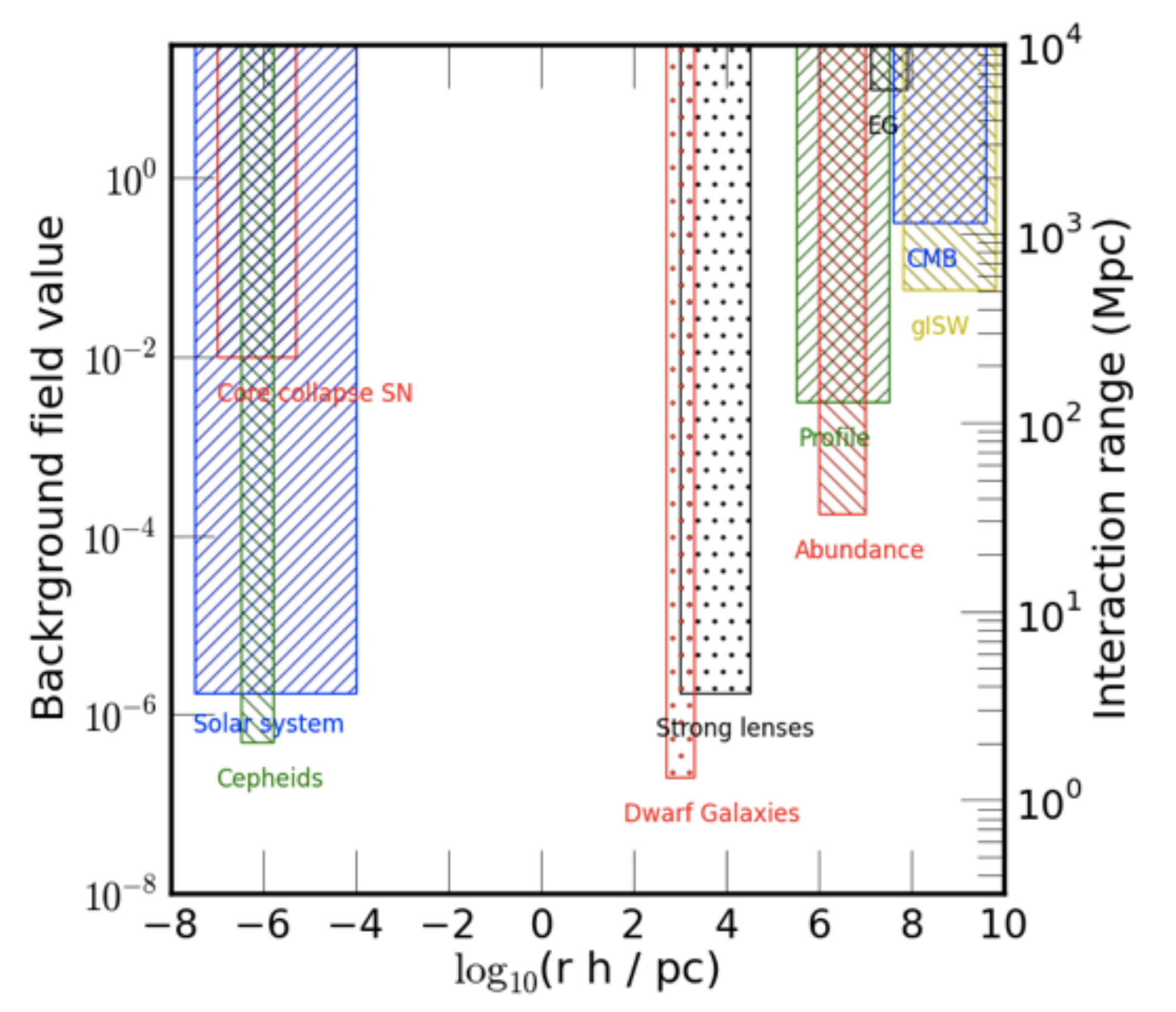}
\end{center}
\label{fig:fR}
\caption{{\bf Left Panel:} Tests of gravity at different length scales. Red
  lines shows observations that probe the sum of metric potentials $\Phi+\Psi$
  via e.g.\ gravitational lensing, while blue lines show dynamical
  measurements that rely on the motions of stars or galaxies or other
  non-relativistic tracers and are sensitive to $\Psi$ alone. Adopted from the
  review by Jain \& Khoury \cite{Jain_Khoury}.  {\bf Right Panel:} 
Astrophysical~\cite{Hu_Sawicki,Jain:2012tn,Vikram:2013uba} and
cosmological~\cite{Song:2007da,Giannantonio:2009gi,Schmidt:2009am} limits on
chameleon theories. The spatial scale on the x-axis gives the range of length
scales probed by particular experiments. The parameter on the y-axis is the
background field value, or the range of the interaction (y-axis label on the
right side) for an $f(R)$ model of the accelerating Universe. The rectangular
regions give the exclusion zone from a particular experiment. All  but the
solar system results have  been obtained in the last 5 years, illustrating the
impressive interplay between theory and experiment in the field. The two
rectangles with dots are meant to indicate preliminary results from ongoing
work. This figure is adapted from Lombriser et al \cite{Lombriser:2011zw}.
}
\end{figure}

Large-scale structure surveys provide a broad set of observables for this
purpose (see Fig.~\ref{fig:fR}).  The abundance, clustering, and motions of
large-scale structure tracers, such as galaxies, clusters, and the
intergalactic medium, can be measured through a variety of methods, such as
photometric and spectroscopic galaxy surveys, X-ray surveys, and small-scale
CMB observations (see Sec.~\ref{sec:clusters}).  Furthermore, the velocities
of tracers such as galaxies can be inferred from their Doppler shifts, which
in turn probe the (non-relativistic) cosmic potential wells.  A fundamentally
different observable is gravitational lensing, measured through the
distortions induced in shapes of background galaxies (Sec.~\ref{sec:WL}).  The
crucial property of gravitational lensing is that it probes all matter,
whether dark or baryonic.  Furthermore, lensing is governed by a spacetime
perturbation different from the perturbation that determines the motion of
non-relativistic bodies such as galaxies.  By comparing velocities (dynamics)
with lensing, one can perform a targeted, model-independent test of gravity
which is largely independent of non-standard cosmological ingredients and
astrophysical systematics.  This is analogous to Solar System tests of
gravity, and corresponds to testing the equivalence of the two Newtonian-gauge
metric potentials.  This and related tests are discussed further in the
Snowmass-2013 paper on Novel Probes of Gravity and Dark Energy
\cite{Jain:2013uma}.

\subsection{Parametrizing Growth}

Within the canonical picture of General Relativity with smooth, late-time,
uncoupled dark energy, the expansion history measured by probes of the recent
Universe completely determines the growth history of structure in the recent
Universe.  However many families of dark energy models lie outside this
picture, and it is important to have methods to detect this and characterize
the deviations in growth, ideally in a model independent manner. In
particular, a desired parametrization of growth should accommodate the
following physics of dark energy and gravity:

\begin{itemize}
\item {\it Clustering} of dark energy, which can arise from a low sound speed,
  $c_s^2\ll1$, common in many high energy physics motivated models such as
  Dirac-Born-Infeld or dilaton theories. Clustering enhances perturbations,
  and while the direct dark energy perturbations are difficult to detect, the
  increased contribution to the gravitational potential through Poisson's
  equation can noticeably affect the matter power spectrum.  Since
  perturbations are only effective when the dark energy equation of state
  deviates from $-1$, this class of models is most visible when there is a
  tracking epoch in the early Universe, where dark energy has an attractor
  solution that may behave like matter or radiation.

\item {\it Early dark energy}, which is the case when dark energy contributes
  to the energy budget at CMB recombination ($a\sim 0.001$, $z\sim 1000$) much
  more than we expect for the cosmological constant case --- that is, much
  more than one part in a billion.  Note that early dark energy budget
  contributions up to 7 orders of magnitude larger (that is, up to $\sim$1\% of the
  total matter plus radiation plus dark energy) are allowed by data. Early
  dark energy affects growth as well, typically suppressing it mildly but over
  a very long cosmological epoch.

\item {\it Couplings} of dark energy, which are interactions between dark energy
  and some other sector. While couplings to Standard Model particles are
  tightly restricted by particle physics data, the possibility remains of
  couplings to dark matter, massive neutrinos, or gravity (this can also be
  viewed as modified gravity, discussed above).
\end{itemize}

To connect these families of models most clearly to the observations of
growth of large scale structure, while remaining general and reasonably
model independent, it is useful to use phenomenological parametrizations
that capture the key impact of these dark energy effects on growth.  This
can be done with a small, remarkably simple set of parameters beyond the
equation of state $w(z)$.

The {\it growth index formalism} provides a general and very
simple-to-implement parametrization that can alert that the growth data is not
consistently following the expectations from the expansion (distance) data.
We define a fitting function for $g(a)$, the linear growth factor divided by
$a$ (i.e.\ with the early matter scaling divided out) in terms of a single
free parameter $\gamma$
\begin{equation}
g(a) \equiv D(a)/a=\exp\left [{\int_0^a (da'/a')\,[\Omega_M(a')^\gamma-1]}\right ]
\end{equation}
where $\Omega_M(a)$ is the energy density relative to critical at epoch
defined by the scale factor $a$. This formula provides an accurate (0.1\%
level) approximation for a wide variety of models
\cite{wangsteinhardt,linder05,lindercahn07}.  Within the canonical picture of
gravity, $\gamma=0.55$, almost independently of $w(z)$.  This separates
out the expansion history as given by $\Omega_M(a)$ from extra growth effects
parametrized by deviations of $\gamma$ from 0.55.  If deviations are detected,
in the value, scale-, or redshift-dependence of $\gamma$, this gives an alert
to check noncanonical models, and in particular modifications of gravity
(where $\gamma$ is closely related to the modified Newton's constant $G_{\rm
  matter}$ introduced below).

Additional parametrizations can shed light on physics behind acceleration. For
example, deviations from a purely matter dominated high redshift growth can be
parametrized by a growth calibration factor $g_\star$, entering the growth
factor as $g=g_\star\exp\left [{\int_0^a
    (da'/a')\,[\Omega_M(a')^{0.55}-1]}\right ]$.  Note that this form
separates out the early time behavior and does not disturb the late time
behavior that was parametrized by $\gamma$ \cite{linder09}.  This test will
pick up early dark energy models.  Moreover, coupled models can affect the
rate of growth --- the tug-of-war between gravitational attraction and the
stretching apart due to accelerated expansion --- at any redshift.  This can be
treated in the growth factor as \cite{linder13,amendola08}
$g=g_\star(f_\infty)\exp\left [{\int_0^a
    (da'/a')\,[f_\infty\Omega_M(a')^{0.55}-1]}\right ]$.  Again, this
preserves the late time value of $\gamma$, keeping that as a distinct alert.
Moreover, since some observations such as redshift space distortions are
directly sensitive to the growth rate, one can measure deviations of
$f_\infty$ from unity fairly directly.

Each of these parametrizations serves as an alert that new physics is in
play, and moreover they identify the region where the new physics enters
--- deviation from expansion predictions, at early times, or couplings.
Once a deviation is detected, the analysis would then concentrate on
more specific models.  For clustered dark energy (evident from a
scale-dependent $\gamma(k)$) for example, one would then introduce models
with a sound speed $c_s$ deviating from the speed of light.

A similar procedure can be applied to the gravitational sector.  Modifications
of gravity will modify how the nonrelativistic and relativistic gravitational
potentials, $\Phi$ and $\Psi$ (which govern the motion of matter and of light,
respectively), are sourced and evolve.  Scalar metric perturbations around a
Friedmann-Robertson-Walker background in the conformal Newtonian gauge are
given by the following spatial metric
\begin{equation}
\label{metric}
ds^2=-a^2(\tau)\left[\left(1+2\Psi\right)d\tau^2-\left(1-2\Phi\right) d\vec{x}^2\right] \ ,
\end{equation}
where $\tau$ and $x$ are the conformal time and distance, respectively.  Using
a model independent parametrization closely tied to the observations, one can
modify the Poisson equations relating matter growth $\delta(a)$ to the
potentials $\Phi$ and $\Psi$.  For example,
\begin{eqnarray}
\nabla^2 \Psi &=& 4\pi G_N a^2\delta\rho \times G_{\rm matter}  \\[0.2cm]
\nabla^2(\Phi+\Psi) &=& 8\pi G_N a^2\delta\rho \times G_{\rm light} \ .
\label{eq:grav_pot}
\end{eqnarray}
Any deviations of dimensionless numbers $G_{\rm matter}$ or $G_{\rm light}$
from unity alerts us to possible modifications of General Relativity.  The
scale- and time-dependence of these parameters can be modeled with independent
$(z, k)$ bins \cite{daniellinder13}, eigenmodes \cite{hojjati13062546}, or
well behaved functional forms
\cite{Bean:2010zq,zhao11091846,Dossett:2011tn,silvestri13021193}.

\section{Cosmological Probes Sensitive to Growth}
\label{sec:probes}

We now proceed to describe how several of the most promising types of
cosmological measurements --- clustering of galaxies in spectroscopic surveys,
counts of galaxy clusters, and weak gravitational lensing --- probe the
growth of structure.

\subsection{Clustering in Spectroscopic Surveys}\label{sec:rsd}

As a natural consequence of large spectroscopic BAO programs such as
the Baryon Oscillation Spectroscopic Survey (BOSS) \cite{dawson13a},
clustering in the density field is sampled at high fidelity in three dimensions
over a wide redshift range.

The galaxies and quasars observed in spectroscopic surveys are biased tracers
of underlying structure, leading to degeneracy between the amplitude
($\sigma^2_R(a)$) of matter fluctuations and biasing parameters.  This
degeneracy complicates the extraction of the growth function from the
isotropic power spectrum derived from clustering of cosmic sources.  However,
spectroscopic surveys encode additional information about the velocity field
arising from gravitational collapse by separately measuring the power spectrum
along and perpendicular to the line of sight.  Because the matter distribution
directly determines the velocity field, these observations can be used to
break the degeneracy between bias and $\sigma^2_R(a)$.  Measurements of the
velocity field help differentiate between the effect of dark energy and
modified gravity as the source of the accelerating Universe through
measurements of Redshift-Space Distortions (RSD) \cite{kaiser87a}.  RSD were
identified by the recent ``Rocky III'' report as ``among the most powerful
ways of addressing whether the acceleration is caused by dark energy or
modified gravity,'' as well as a tool to increase the dark energy Figure of
Merit from spectroscopic surveys.

RSD arise because the gravitational pull of matter over-densities causes
velocity deviations from the smooth Hubble flow expansion of the Universe.
These peculiar velocities are imprinted in galaxy redshift surveys in which
recessional velocity is used as the line-of-sight coordinate for galaxy
positions, leading to an apparent compression of radial clustering relative to
transverse clustering on large spatial scales (a few tens of Mpc). On smaller
scales (a few Mpc), one additionally observes the ``finger-of-God'' elongation
\cite{Jackson_FoG} due to random velocities of galaxies in a cluster.  The
resulting anisotropy in the clustering of galaxies is correlated with the
speed at which structure grows; deviations from GR causing slower or faster
growth give smaller or larger anisotropic distortions in the observed
redshift-space clustering.  RSD are sensitive to the rate of change of the
amplitude of clustering, $f\sigma_8(a) = d\sigma_8(a)/d\ln a$, where
$a=(1+z)^{-1}$ is the dimensionless cosmic expansion factor and $f\equiv f(a)$
has been defined in Eq.~(\ref{eq:f}).  Because RSD measurements are sensitive
to the product of the growth rate and the amplitude of matter fluctuations, a
wide range in redshift coverage is essential to constrain the evolution in
clustering amplitude and directly probe GR.  On the other hand, if one were to
assume a $\Lambda$CDM model where GR correctly explains gravitational
collapse, the growth rate can be predicted to high precision and RSD results
can be used to constrain $\sigma_8(a)$, thereby providing insight into other
fundamental physics such as neutrino masses.

\begin{figure}[t]
\centering
\includegraphics[height=0.35\textwidth]{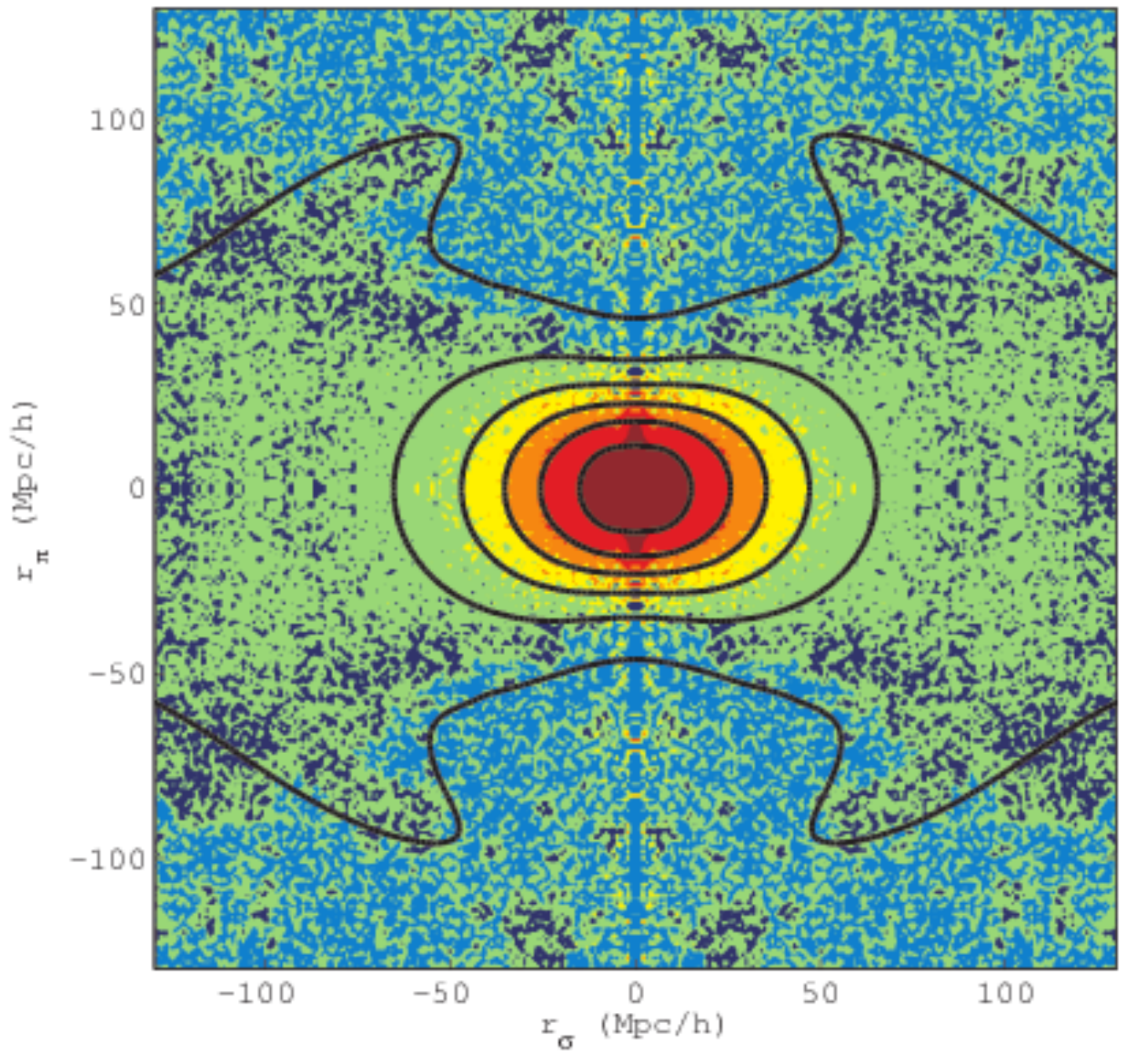}
\includegraphics[height=0.37\textwidth]{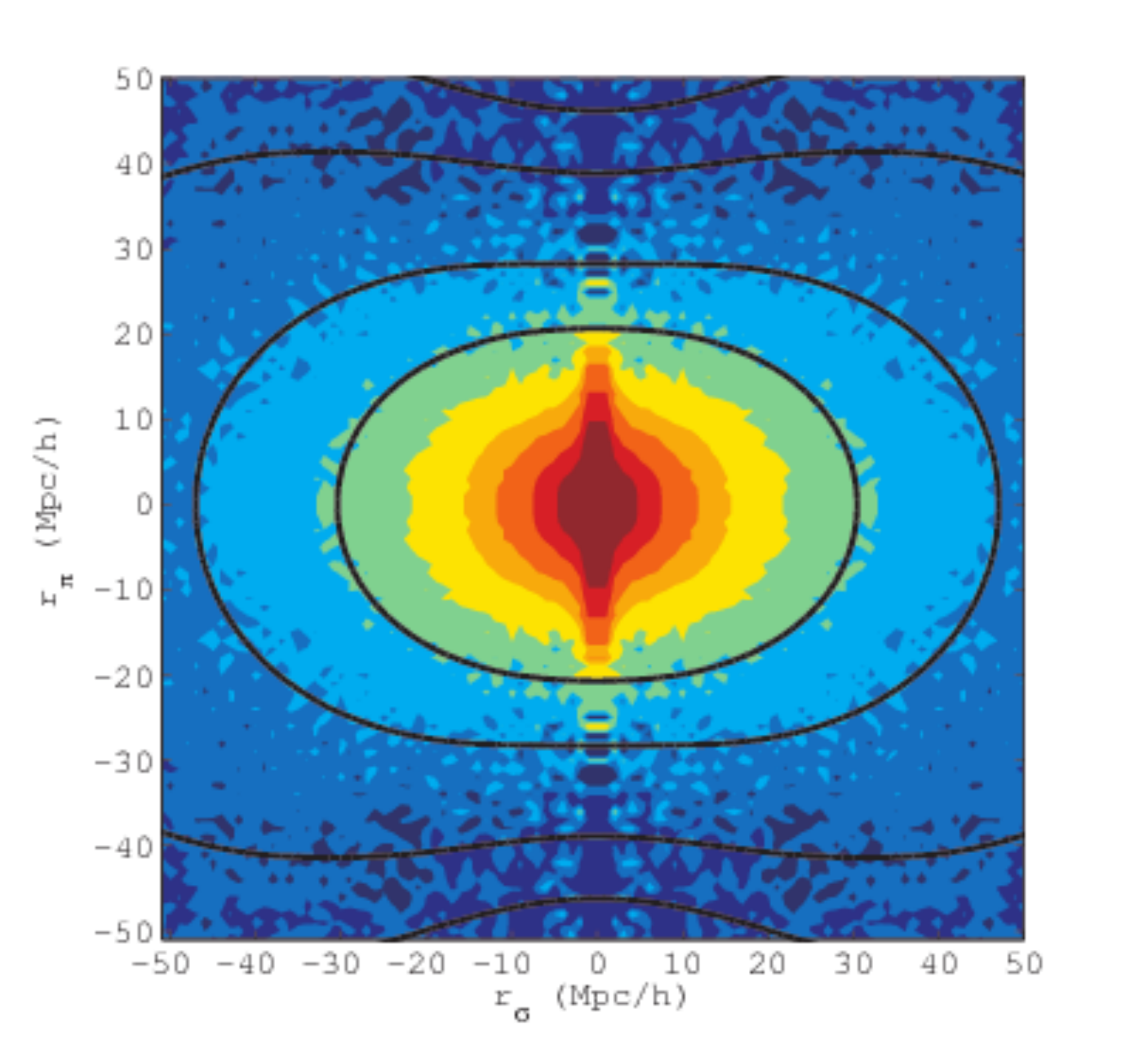}
\caption{{\bf Left panel}: Two-dimensional correlation function of BOSS
  galaxies (color) compared with the best fit model (black lines).  Contours
  of equal $\xi$ are shown at [0.6, 0.2, 0.1, 0.05, 0.02, 0].  {\bf Right
    panel}: Smaller-scale two-dimensional clustering with model contours at
  [0.14, 0.05, 0.01, 0].  Figures adopted from Ref.~\cite{reid12a}.
}
\label{fig:butterfly}
\end{figure}

Some of the earlier measurements of RSD were obtained by the Two Degree Field
Galaxy Redshift Survey \cite{Percival:2004fs}, the Vimos-VLT Deep Survey
\cite{Guzzo:2008ac}, the 2SLAQ survey \cite{daAngela:2006mf}, the SDSS-II
\cite{Cabre:2008sz,Samushia:2011cs,Raccanelli:2012gt}, and WiggleZ \cite{Blake:2011rj}.  More
recently, Refs.~\cite{reid12a} and \cite{samushia13a} presented the first
measurements and cosmological interpretation of RSD in the two-year BOSS
galaxy sample \cite{ahn12a}.  Figure \ref{fig:butterfly} shows their
measurement of the correlation function in terms of the line-of-sight
separation and transverse separation.  The central ``squashing'' evident in
the left panel is due to structure growth.  The right panel of the same figure
shows the clustering signal on smaller scales; the ``finger-of-God''
elongation effect from velocities on small scales is visible for small
transverse separations but unimportant on the scales of RSD analyzed here.
With these results, BOSS constrains the parameter combination
$f\sigma_8(a)=0.43\pm 0.07$ at the mean redshift of the sources $z=0.57$ (or
$a=0.64$), improving to $f\sigma_8(a)=0.415 \pm 0.034$ if assuming a
$\Lambda$CDM expansion history.

\subsubsection{Prospects for Future Measurements}

The Extended Baryon Oscillation Spectroscopic Survey (eBOSS) is the
cosmological survey within SDSS-IV, a six year program that will begin in
August 2014.  eBOSS will provide the first percent-level distance measurements
with BAO in the redshift range $1<z<2$, when cosmic expansion transitioned
from deceleration to acceleration.  The targets for eBOSS spectroscopy will
consist of: Luminous Red Galaxies (LRGs: $0.6<z<0.8$) at a density of 50
deg$^{-2}$, Emission Line Galaxies (ELGs: $0.6<z<1.0$) at a density of 180
deg$^{-2}$, ``clustering'' quasars to directly trace large-scale structure
($1<z<2.2$) at a density of 90 deg$^{-2}$, re-observations of faint BOSS
Lyman-$\alpha$ quasars ($2.2<z<3.5$) at a density of 8 deg$^{-2}$, and new
Lyman-$\alpha$ quasars ($2.2<z<3.5$) at a density of 12 deg$^{-2}$.

The extended redshift range of the combined eBOSS and BOSS measurements will
significantly reduce the degeneracy between $f(a)$ and $\sigma_8(a)$.  The wide
redshift range of the eBOSS tracers will allow a separation of the evolution
of structure growth from the amplitude of clustering and provide new
constraints on GR through RSD analyses, or provide tight constraints on
$\sigma_8$ in the assumption of $\Lambda$CDM.       The projections on RSD
  constraints from eBOSS are computed from a Fisher matrix formalism assuming
  measurements of large-scale modes with wavelengths up to $k_{\rm
    max}=0.2\ h\rm\ Mpc^{-1}$.  The expected 68\% confidence constraints on
the growth of structure, parametrized as $f\sigma_8(a)$ and measured from RSD,
are $\sigma_{f\sigma_8(a)}/f\sigma_8(a) = $ 0.029, 0.035, and 0.036 for the
LRG, ELG, and quasar programs respectively.

Prime Focus Spectrograph (PFS) will be a powerful spectroscopic survey of
faint emission galaxies because of its large multiplex gain and the 8.2 meter
aperture of the Subaru telescope \cite{Ellis:2012rn}. The extended wavelength
coverage provided by the red and near-infrared spectrograph arms (650 -– 1260
nm) will permit a survey of about $2\times 10^6$ [O II] emission-line galaxies
extending over the redshift range $0.8 < z < 2.4$. As large-scale structure is
still in the linear regime at high redshift, such a deep survey will give
detailed new information on the cosmological parameters as well as the growth
of structure. Multi-color data planned to arrive from the Hyper Suprime Cam
(HSC) imager will be used to select target galaxies for spectroscopy.  The
proposed PFS cosmology survey will consist of 100 nights of observations
surveying over 1400 sq. deg., sampling galaxies within a comoving volume of
9(Gpc/$h$)$^3$. This will complement the lower redshift survey being
undertaken by the BOSS collaboration.  Apart from accurately measuring the
dark energy parameters and being sensitive to the presence of early dark
energy from the geometrical BAO measurements, PFS will measure the RSD out to
redshift $z=2.4$, and provide the measurement of $f(a)$ to 6\% accuracy in
each of six bins spanning its redshift range \cite{Ellis:2012rn}. These PFS
measurements of the large scale galaxy distribution can be combined with
complementary weak lensing information from the HSC survey in order to improve
the growth constraints and reduce uncertainties arising from galaxy bias and
nonlinearities that are otherwise major sources of systematic error in
spectroscopic surveys.

Dark Energy Spectroscopic Instrument (DESI) will be the largest and most
powerful ground-based spectroscopic survey.  DESI will provide a comprehensive
survey of at least 14,000 deg$^2$ with an order of magnitude more
spectroscopic galaxies and quasars than obtained in BOSS and eBOSS combined.
As with eBOSS, the primary targets will be derived from LRG, ELG, and quasars
selected from imaging data.  The redshift ranges will be refined to probe the
$z>0.6$ epochs at higher resolution than eBOSS: LRG targets will cover
$0.6<z<1.0$, ELG targets will cover $0.6<z<1.5$, and quasar targets will cover
$1<z<3.5$.  In projecting RSD constraints, we assume a 14,000 deg$^2$ survey
with densities 1325 deg$^{-2}$, 300 deg$^{-2}$ and 176 deg$^{-2}$ for ELG,
LRG, and quasars, respectively.  Because the number density of galaxies is
sufficient to provide RSD constraints in finely binned samples ranging from
$0.1<z<1.8$, we do not report all of the projections here.  The expected 68\%
confidence constraints from RSD, following the same assumptions as above, are
shown in Figure~\ref{fig:killer}.  The precision in each bin ($\Delta z=0.1$)
is better than 2\% from $0.4<z<1.5$, while the aggregate accuracy from the
combination of all three tracers is $\sigma_{f\sigma_8(a)}/f\sigma_8(a) =
0.0035$.

\subsubsection{Challenges in RSD Constraints}

The RSD projections are calculated using the methodology of \cite{white09a},
which assumes that the shape of the power spectrum and the cosmological
distance-redshift relationship are known perfectly.  While this method
produces predictions that are thus independent of additional data sets,
marginalization over the power
spectrum shape and distance-redshift relationship can potentially degrade the
growth constraints. For BOSS, the remaining uncertainty on the shape of the
power spectrum is negligible, given a prior from the cosmic microwave
background.  Imperfect knowledge of the cosmological distance-redshift
relationship induces additional anisotropy in the observed galaxy correlations
via the Alcock-Paczynski effect \cite{AP,Matsubara:1996nf,Ballinger:1996cd}
that is partially degenerate with the RSD-induced anisotropy.  The
Alcock-Paczynski effect depends on the product $D_A(a)H(a)$, where $D_A$ is
the angular diameter distance and $H$ is the Hubble parameter.  However,
allowing for an arbitrary value of $D_A(a)H(a)$ degraded $f\sigma_8$
constraints by a factor of two in BOSS.  Marginalizing over additional
parameters in nonlinear galaxy biasing and small-scale Finger-of-God
velocities may further slightly degrade the errors (by 10\% for BOSS DR9
analysis).


RSD have, until recently, been modeled using a simplistic separation of
density and velocity correlations. This separation is known to be inaccurate
especially at smaller scales, where non-linear growth ($\delta >1$) couples
density and velocity modes at different scales. It is going to be necessary
to simulate and understand these correlations to sufficient precision to
avoid systematic errors, from imprecise modeling, that  degrade the
reconstruction of the growth of structure from RSD observations
\cite{Kwan:2011hr}.

Finally, the projections assume that spectroscopic large-scale clustering
measurements will be limited by statistical errors.  This requires stringent
control of systematic errors that can modulate the data on varying scales,
such as the impact of stellar contamination and dust extinction on target
selection efficiency, variations in seeing that alter target selection and
redshift success, and so on.  These systematics have already been extensively
studied within BOSS \cite{ross11a, ross12b}, and the greater volume and
greater statistical power at large scales from eBOSS and DESI will place new demands on
homogeneity of the target samples.  Lessons from BOSS are being applied to
target selection in eBOSS\@.  Similarly, better understanding of these
systematics learned during eBOSS will provide important information for
preparation of target selection in DESI\@.

Given the potential for providing dark energy measurements, constraints on
fundamental neutrino and inflation physics, and cluster redshifts, velocity
dispersions, and calibration of photometric redshifts for large imaging
programs, it is clear that wide-field optical spectroscopy will play a central
role in cosmology well after the completion of DESI.  At this time, it is
impossible to estimate the exact details of such a program, but detector
technology, full integration of robotic fiber positioners into large surveys,
and likely availability of large telescopes paint a clear path for
ground-based spectroscopic surveys beyond DESI.  One can imagine an order of
magnitude increase in the number of fibers per field of view, sensitivity at
least one magnitude fainter than that of DESI, and wavelength coverage
extending in the near infrared.  The  \WFIRST\ 
and \Euclid\ missions both incorporate near-IR, slitless spectroscopic surveys
that are expected to detect tens of millions of emission line galaxy redshifts
in the range $0.5<z<2$.

Precise modeling of target source populations, their respective number
densities and redshifts is beyond the scope of this document. In addition, the
power of these surveys will likely only be fully realized when the theoretical
models of structure formation adequately describe the velocity field measured
in non-linear and mildly non-linear regimes.  Theoretical developments in the
growth of structure are discussed in more detail in Sec.~\ref{sec:sim}.

\subsubsection{Other Uses of Clustering to Probe Growth}

Along with the geometric information from baryon acoustic oscillations, the
growth information from the RSD is thought to provide the most promising and
reliable method that uses clustering of galaxies to measure dark energy
properties. However, growth can be probed using several other methods that use
either photometric or spectroscopic galaxy surveys, and here we cover them
briefly. 

The broadband power spectrum of galaxies or other tracers of the large-scale
structure, $P(k, a)$, can be measured to excellent accuracy over several
decades in $k$. On linear scales, the power spectrum is proportional to
$D(a)^2$, and hence it directly probes the growth of structure. Unfortunately,
the power spectrum is also proportional to the bias of the tracer objects, and
this bias is typically {\it also} time and scale-dependent, albeit in a way
that often has to be extracted from the data itself. It is therefore
challenging to obtain accurate constraints on the growth of structure from the
broadband $P(k, a)$ measurements alone. On the other hand, combining the
broadband power measurements that are sensitive to bias with weak lensing
measurements that are not can be used to break this bias-growth
degeneracy. This is one of the manifestations of the powerful synergy between
the spectroscopic and photometric surveys.

The cross-correlation between the galaxy density field and the hot and cold
spots in the CMB anisotropy maps is directly sensitive to the Integrated
Sachs-Wolfe (ISW) effect, and thus probes the decay of the gravitational
potential due to the presence of dark energy at late times. These measurements
produced independent evidence for dark energy and have achieved increased
accuracy over the years
(e.g.\ \cite{Boughn:2003yz,Giannantonio:2008zi,Ho:2008bz}). However the
largest signal available from the cross-correlation corresponds to about
10-$\sigma$ detection of the effects of dark energy via the ISW effect
\cite{Hu:2004yd}, making it a probe with relatively modest prospects.

Cosmic magnification, discussed in more detail in the Snowmass-2013 paper on
Cross-Correlations and Joint Analyses \cite{Rhodes:2013tma}, induces
additional spatial correlations between the density, luminosity, and size of
objects due to the bending of light by structures located between those
objects and the observer.  The full potential of magnification measurements to
probe dark energy is only beginning to be explored.

\subsection{Cluster Abundances}\label{sec:clusters}

\subsubsection{Clusters Abundances as a Probe of Fundamental Physics}

Galaxy clusters are the most massive gravitationally bound structures in the
Universe.  As with other dynamical probes, their primary importance in the
context of dark energy is their complementarity to geometric probes,
i.e. their ability to distinguish between modified gravity and dark energy
models with degenerate expansion histories.  For a complete review, we refer
the reader to Refs.\ \cite{allenetal11} and \cite{weinbergetal12}.

The basic physics behind cluster abundances as a cosmological probe are
conceptually simple.  Clusters form the gravitational collapse of density
fluctuations.  Prior to collapse, the growth of these fluctuations is linear,
so that the matter density contrast $\delta=\delta\rho/\rho$ at {\it any}
spatial scale evolves with the same growth factor $D(a)$, $\delta(a)\propto
D(a)$, where the linear growth factor $D(a)$ can be evaluated exactly for a
given cosmological model following Eq.~(\ref{eq:growth}).

At some critical threshold $\delta_c$, the perturbation undergoes
gravitational collapse.  Consequently, the probability of forming a halo --- a
collapsed object --- is equivalent to the probability that $\delta \geq
\delta_c$.  Assuming Gaussian random initial conditions, one finds
then that the number of collapsed objects $N$ per unit mass $dM$ and
comoving volume element $dV$ is
\begin{equation}
{dN\over dMdV} = 
F(\sigma)\frac{\rho_M}{M}
\frac{d\ln\sigma^{-1}}{dM} \, ,
\end{equation}
where $\rho_M$ is the matter density in the Universe, $\sigma^2$ is the
variance of the density perturbations evaluated at some mass scale $M$ (or,
equivalently, spatial scale $R$ where $M=(4\pi/3) R^3 \rho_M$). Here
$F(\sigma)$ corresponds to the fraction of mass in collapse object;
in the Press--Schechter argument \cite{PressSchechter},
\begin{equation}
F(\sigma) = \frac{1}{\sqrt{2\pi\sigma^2}}
\exp{\left (-\frac{1}{2}\frac{\delta_c^2}{\sigma^2}\right )} \, ,
\end{equation}
 while cosmological N-body simulations have been used to calibrate $F(\sigma)$
 to higher precision beyond the simple Press--Schechter formula.

It is precisely this dependence of the number of galaxy clusters on the
variance of the {\it linear} density field that allows us to utilize galaxy
clusters to constrain the growth of structure.  In particular, the late-time
variance of the the linear density field $\sigma^2\equiv \sigma^2(a)$ is
related to the variance at some initial scale factor $a_0$, $\sigma_0^2$, via
the linear growth function, $\sigma^2 = [D^2(a)/D^2(a_0)]\sigma_0^2$, which
makes the abundance of galaxy clusters explicitly dependent on the growth
history of the Universe.


\begin{figure*}
\begin{center}
\includegraphics[width=0.45\textwidth]{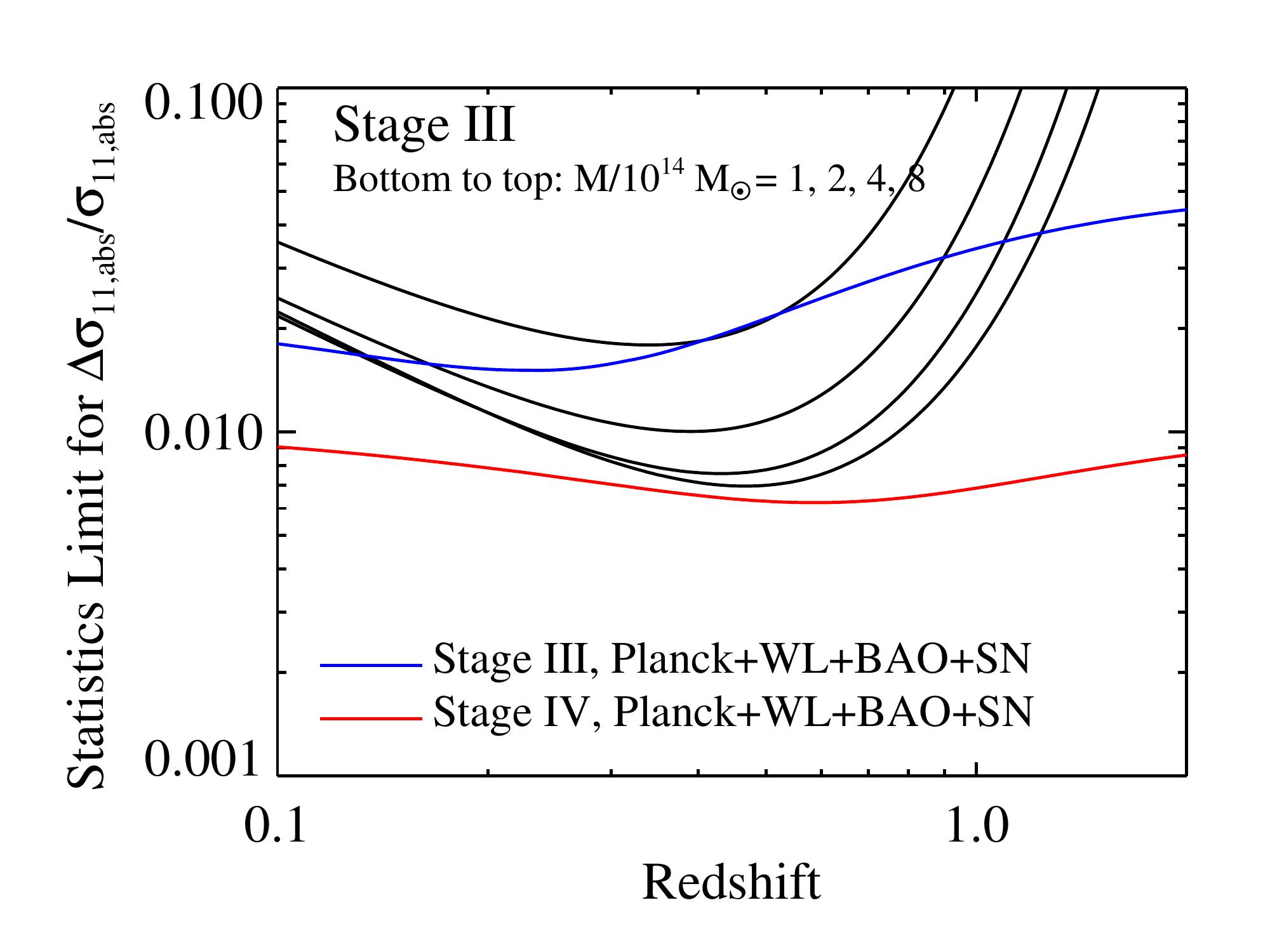} 
\includegraphics[width=0.45\textwidth]{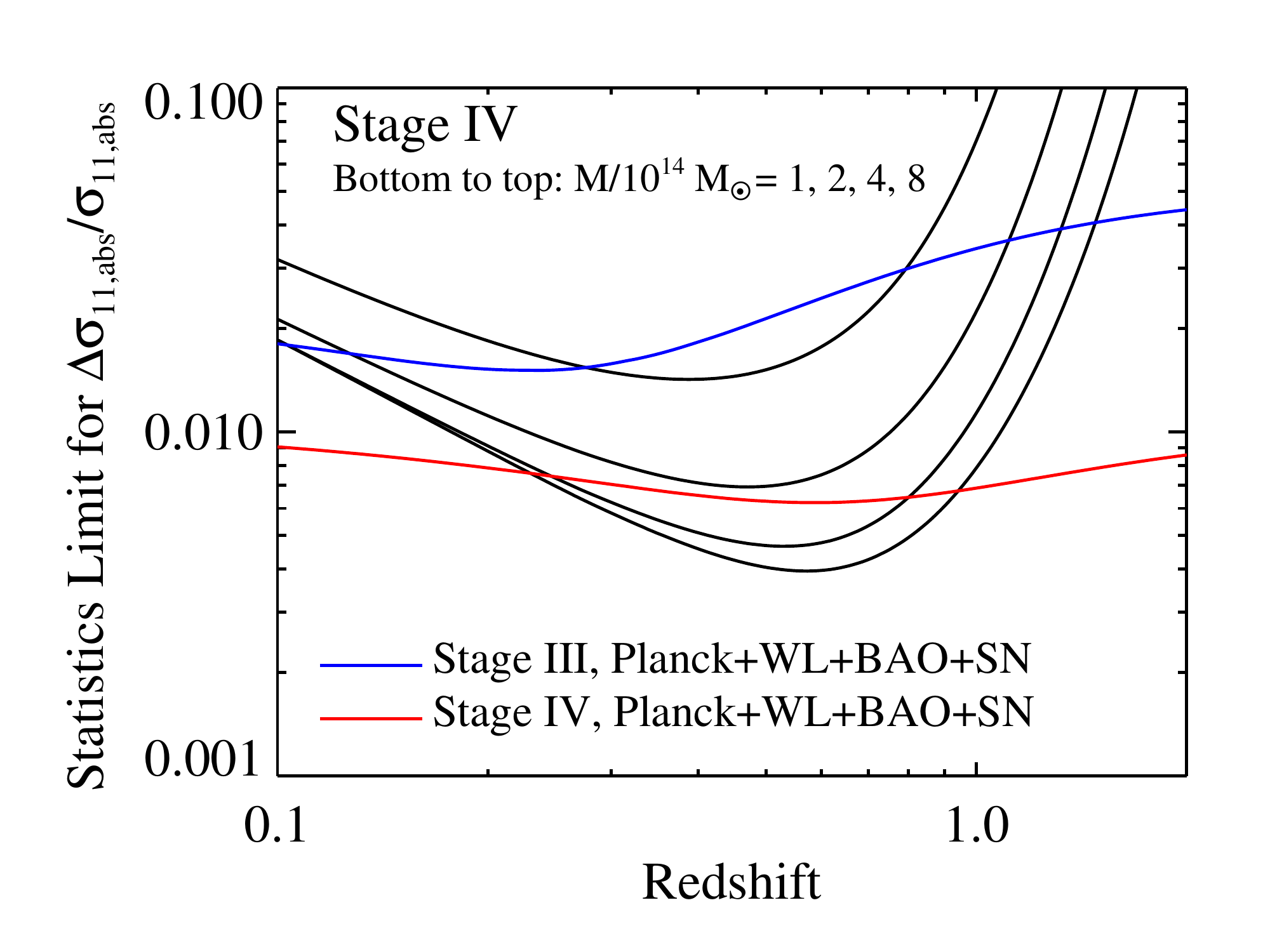}
\end{center}
\vspace{-0.6truecm}
\caption{Statistical error on $\sigabs(z)$ --- or equivalently the growth
  function $D(a)$ --- derived using galaxy clusters in redshift bins of width
  $z\pm 0.05$ for two fiducial $10^4\ \deg^2$ surveys: {\bf Left panel}: a
  Stage III survey with shape noise appropriate for ground-based imaging
  (source surface density $10\ \mathrm{gals}/\mathrm{arcmin}^2$, shape noise
  $\sigma_e=0.4$); {\bf Right panel}: a Stage IV survey with shape noise
  appropriate for space-based imaging ($30\ \mathrm{gals}/\mathrm{arcmin}^2$,
  $\sigma_e=0.3$).  Each solid black line corresponds to a different mass
  threshold, as labeled.  The blue and red curves are the corresponding
  constraints from Stage III (blue) and Stage IV (red) weak lensing +
  supernova + BAO + \Planck\ experiments.  Figure taken from
  Ref.~\cite{weinbergetal12}.  }
\label{fig:s11}
\end{figure*}


A cluster abundance experiment is conceptually very simple: one first
identifies galaxy clusters --- e.g.\ as cluster of galaxies in the optical, as
extended sources in the X-rays, or as Sunyaev-Zel'dovich (SZ) sources (cold
spots in the CMB) at millimeter wavelengths --- and then one needs to
determine the corresponding cluster masses.  As noted below, future surveys
will almost certainly rely on weak lensing mass calibration to estimate
cluster masses.  Figure~\ref{fig:s11}, taken from Ref.~\cite{weinbergetal12}, shows
forecasted constraints on the amplitude of matter fluctuation $\sigabs$ for
two fiducial $10^4\ \deg^2$ surveys, assuming
$10\ \mathrm{gals}/\mathrm{arcmin}^2$ and a shape noise appropriate for
ground-based observations ($\sigma_e=0.4$), and for a similar survey with
$30\ \mathrm{gals}/\mathrm{arcmin}^2$ and a shape noise appropriate for
space-based observations ($\sigma_e=0.3$).  These forecasts are compared to
the predictions for Stage III (blue curve) and Stage IV (red curve)
\Planck+weak lensing+supernova+BAO experiments.  We see that galaxy clusters
are statistically competitive with and often better than other probes,
highlighting their complementarity as a discriminant between dark energy and
modified gravity models.

Galaxy clusters can probe dark energy in other ways as well, most
notably by comparing cluster mass estimates from weak lensing and dynamical
methods such as galaxy velocity dispersions; see Sec.~\ref{sec:DE-MG}.  In
addition, because the growth of structure is also impacted by non-zero
neutrino mass, galaxy cluster abundances can provide competitive constraints
on the sum of neutrino masses, further enhancing their value as a tool for
fundamental physics (e.g.\
\cite{mantzetal10c,houetal12,burenin13,rozoetal13}).


\subsubsection{Systematics: Current Limitations and Future Prospects}

Galaxy clusters can be identified with optical, mm, or X-ray data.  Regardless
of how the clusters are identified, the cosmological utility of cluster
samples is always limited by our ability to estimate the corresponding cluster
masses. Roughly speaking, CMB+geometric probes predict the amplitude of matter
fluctuations as a function of redshift with $\approx 3\%$ (0.9\%) precision
for Stage III (IV) dark energy experiments.  To achieve comparable levels of
precision using galaxy clusters, we must be able to measure cluster masses
with $\approx 5\%$ (2\%) precision \cite{weinbergetal12}.

While this level of precision is significantly better than what has been
achieved to date, there are good reasons to believe that it can be achieved in
the near future.  Specifically, most cosmological work to date has relied on
hydrostatic X-ray mass estimates, which are subject both to departures from
hydrostatic equilibrium
(e.g.\ \cite{nagaietal07a,lauetal09,battagliaetal11,rasiaetal12}) and X-ray
calibration uncertainties \cite{nevalainenetal10,tsujimotoetal11}.  Future
work, however, will rely on weak lensing mass calibration, which entirely
bypasses the aforementioned systematics at the expense of new, better
controlled systematics.

The two primary sources of systematic errors for weak lensing mass calibration
are shear biases, i.e. systematic uncertainties in our estimates of the
gravitational shear, and systematic errors in the redshift distribution of the
photometric lensing sources.  Where the systematic floor of these type of
measurements ultimately remains to be seen, with the most recent analyses
suggesting that $\approx 7\%$ mass calibration has been achieved.  We caution,
however, that $\approx 20\%$ systematic offsets between different groups
remain \cite{applegateetal12}.

The reduction of shear and photometric redshift systematics is the thrust of
ongoing investigations.  Shear estimation methods are being tested and
improved upon via extensive simulation tests
\cite{STEP1,STEP2,Great10,kitchingetal12b}, and the possibility of
self-calibrating systematics from joint shear and magnification analyses of
the weak lensing signal has been noted
\cite{vallinottoetal10,rozoschmidt10,huffgraves11,schmidtetal11}.  In
addition, the use of spectroscopic sources for weak lensing mass calibration
\cite{couponetal13} entirely bypasses both sources of systematic errors.
Similarly, photometric redshift errors have been the focus of several recent
theoretical works aimed specifically at understanding how to minimize this
source of systematic uncertainty
(e.g.\ \cite{hearinetal10,cunhaetal12,hutereretal13}).  Alternatively, as
discussed in the Snowmass-2013 white paper on the Spectroscopic Needs for
Imaging Dark Energy Experiments \cite{Newman:2013pma}, cross-calibration
methods may provide an effective alternative to photometric redshift biases
\cite{newman08,mcquinnwhite13}.  In short, {\it there are very good reasons to
  believe that the systematic floor in current weak lensing measurements will
  be significantly reduced in the future.}

Systematic errors in shear measurements tend to be less critical for cluster
abundance work than for cosmic shear work, partly because of the existence of
a preferred orientation {\it a priori} (we are interested in tangential
shear), partly because of circular averaging of the shear (which removes
systematics that fluctuate on scales larger than a galaxy cluster), and partly
because the weak lensing signal of clusters is large relative to the typical
shear signal.  The impact of photometric redshifts on both data sets, however, is fairly
comparable.  Overall, one can fairly generically state that experiments that
control shear systematics at a level that enables cosmic shear experiments
also automatically enable cluster weak lensing mass calibration.

There are, however, additional (currently sub-dominant) sources of systematics
that can impact galaxy clusters.  Key amongst these is the calibration not
only of the mean relation between cluster observables (optical, X-ray, or mm
signals) and cluster mass, but also the scatter (shape and amplitude) about
the mean.  Estimates (e.g.\ \cite{wuetal10}) suggest that $\approx 5\%$
calibration of this scatter --- which is achievable today --- is sufficient
for near future experiments (e.g.\ DES, HSC, PanSTARRS), but this source of
error is likely to become significant for Stage IV surveys such as Large
Synoptic Survey Telescope (LSST), \Euclid, and \WFIRST, and certainly for any
putative Stage V experiment.  Such calibrations should be achievable with high
resolution X-ray imaging using high quality mass proxies like $M_{gas}$ or
$Y_X$.  Note that these proxies will themselves be calibrated via weak
lensing, so the hydrostatic bias noted above for X-ray mass calibration is
irrelevant in this context.

In addition, cluster centering remains an important systematic in optical
and/or low resolution experiments (e.g.\ \Planck).  Specifically, weak lensing
mass calibration requires we measure the tangential shear of background
galaxies centered on galaxy clusters, but selecting the center of a galaxy
cluster is not always trivial.  This systematic can either be self-calibrated
\cite{oguritakada10}, or it may be calibrated with high resolution X-ray/mm
follow-up of small sub-samples optical galaxy clusters.  Note that optical
cluster detection is still highly desirable, as optical observations benefit
from a lower mass detection threshold than X-ray/mm over a large redshift
range, which in turn result in improved statistical constraints.  Thus, the
combination of optical with X-ray and mm data is clearly superior than either
data set alone.

{\it The synergistic nature of multi-wavelength cluster cosmology will
  necessarily play a key role in future cluster abundance experiments.}
Clusters are fortunate in that they can be studied across the electromagnetic
spectrum, and consistency between all measurements provide critical
self-consistency constraints that can ferret out hitherto undetected
systematics \cite{rozoetal12d}.  Moreover, multi-wavelength cluster
abundance studies can further improve cosmological constraints relative to
what can be achieved with single wavelength measurements
\cite{wuetal10,cunha09}.  Consequently, a balanced multi-wavelength approach
will be critical to the success of cluster cosmology over the next 10-20
years.

One final key prospect with galaxy clusters remains, that of self-calibration.
That is, the cluster-clustering signal is itself an observable that one can
use to calibrate cluster masses, and which is insensitive to all of the above
systematic effects.  In general, self-calibration does result in some
degradation of cosmological information relative to systematics-free weak
lensing measurements, but such loss decreases with a decreasing mass
threshold.  Provided one can reach low cluster masses, e.g.\ in the optical,
self-calibration is an attractive option.  Indeed, multiple studies have found
that self-calibration of galaxy clusters are capable of placing cosmological
constraints that are comparable to those from cosmic shear analysis in the
absence of systematics \cite{cunhaetal09,oguritakada10}.


\subsubsection{Cluster Wish Lists}

The minimum necessary data for Stage IV cluster experiments should be
immediately available from LSST, \Euclid, and \WFIRST\ in the optical/IR, eRosita
in the X-rays, and \Planck, and the new generation South Pole Telescope (SPT)
and Atacama Cosmology Telescope (ACT) experiments.  We emphasize that scatter
calibration --- regardless of the origin of the clusters --- will require high
resolution X-ray imaging, so large follow-up programs with existing
instruments (Chandra, XMM) is a must.  The commissioning of new, more sensitive
  high resolution X-ray satellites is particularly important for studying the
  lower mass, higher redshift systematics that should dominate the next
  generation of cluster cosmology experiments.  In the mm, next generation
  surveys such as SPT3G and beyond will allow for better mass calibration and
  systematics control of optical surveys at low cluster masses, as well as SZ
  detection of higher redshift systems.  Continuing improvement in mm detector
  technology will remain a fruitful enterprise from the point of view of
  cluster cosmology.

In addition, spectroscopic follow-up of 2+ bright galaxies in galaxy clusters
is highly desirable.  As an example, in its Spectroscopic Identification of
eROSITA Sources (SPIDERS) program, eBOSS will acquire several redshifts per
cluster by observing objects associated with eROSITA x-ray clusters but not
included in the BOSS and eBOSS galaxy clustering samples.  Such a follow-up
program will lead to improved cluster centering, better photometric redshift
performance and calibration, and it would enable testing of modified gravity
models by comparing dynamical to weak lensing masses via correlation methods.
Note that this program could easily be included as part of a spectroscopic
program targeting LRGs, as these type of galaxies dominate the cluster
population, an obvious ``value added'' to spectroscopic BAO surveys, provided
the spectroscopic and cluster surveys overlap.  Similarly, spectroscopic
follow-up of background galaxies at high redshift for BAO studies
(e.g.\ emission-line galaxies) enables spectroscopic weak lensing
measurements, which are insensitive to both shear and photometric redshift
systematics.  These spectroscopic samples also allow for photometric redshift
calibration via cross-correlation methods, which will reduce systematic error
uncertainties in the redshift distributions of photometric source galaxies.
In short, from a cluster perspective, overlap of cluster surveys with
spectroscopic BAO experiments clearly provides a value added to clusters that
is otherwise unavailable.  A quantification of these gains, however, is
difficult, as the value of such measurements will likely depend on the
systematic floor of shear-based weak lensing mass measurements with
photometric sources.

\subsection{Weak Gravitational Lensing} \label{sec:WL}

\subsubsection{Background}

The gravitational bending of light by structures in the Universe distorts or
shears the images of distant galaxies. This distortion allows the
distribution of dark matter and its evolution with time to be measured,
thereby probing the influence of dark energy on the growth of structure.

Within the past decade, {\it weak gravitational lensing} --- slight
distortions of galaxy images due to the bending of the light from distant
galaxies by the intervening large-scale structure --- has become one of the
principal probes of dark matter and dark energy. The weak lensing regime
corresponds to the intervening surface density of matter being much smaller
than some critical value.  While weak lensing around individual massive halos
was measured in the 1990s \cite{Tyson90,Brainerd95}, weak lensing by
large-scale structure was eagerly expected, its signal predicted by theorists
around the same time \cite{Miralda-Escude91,Kaiser92,Jain_Seljak}.  In this
latter regime, the observed galaxies are slightly distorted (roughly at the
1\% level) and one needs a large sample of foreground galaxies in order to
separate the lensing effect from the noise represented by random orientations
of galaxies.

A watershed moment came in the year 2000 when four research groups nearly
simultaneously announced the first detection of weak lensing by large-scale
structure \cite{Bacon_detect,Kaiser_detect,vW_detect,Wittman_detect}. Since
that time, weak lensing has grown into an increasingly accurate and powerful
probe of dark matter and dark energy
\cite{RCS,CTIO,Subaru,VIRMOS-Descart,GEMS,COSMOS,COMBO17,COSMOS-tomo,Huff,DLS,Heymans-CFHTLens}.
Below we first briefly summarize how weak lensing probes dark energy and in
particular the growth of structure; more detailed reviews of the topic are
available in \cite{Refregier:2003ct,Hoekstra:2008db,Huterer:2010hw}.

\subsubsection{Shear Measurements and Cosmological Constraints}

The statistical signal due to gravitational lensing by LSS is termed cosmic
shear. The cosmic shear field at a point in the sky is estimated by locally
averaging the shapes of large numbers of distant galaxies. The primary
statistical measure of the cosmic shear is the shear angular power spectrum,
which is measured as a function of the source-galaxy redshift $z_s$. Additional
information is obtained by measuring the correlations between shears at
different redshifts, which is referred to as 'shear tomography', or between
shears and foreground galaxies --- the ``galaxy-galaxy lensing''.

The principal power of weak lensing comes from the fact that it responds to
all matter, both dark and baryonic, and not just to visible (or, more
generally, baryonic-only) matter like most other probes of the large-scale
structure. Therefore, modeling of the visible-to-dark matter bias, a thorny
and complicated subject, is altogether avoided when using weak lensing.
Simulations of dark matter clustering are becoming increasingly accurate, and
simulation-based predictions that include baryons (which steepen and therefore
affect the dark matter halo density profiles \cite{Rudd:2007zx,Yang:2012qq,Zentner:2012mv})
should be able to reach the accuracy required to model the weak lensing signal
so that modeling errors do not appreciably contribute to the total error
budget. Because of our ability to model its signal accurately, weak lensing
has great intrinsic power to probe dark matter and dark energy in the
Universe.

The other principal reason why weak lensing is powerful comes from the fact that {\it
  galaxy shear is sensitive to both geometry and the growth of
  structure}. Gravitational lensing depends on the geometry (e.g.\ location of
the lens relative to the source and the observer and the mutual distances
involved), while the growth determines how much structure is available at a
given distance to serve as cosmic lenses for light coming from even more
distant galaxies. In particular, in the so-called Limber approximation and
assuming small shear, one can write the two-point correlation function of
shear in harmonic space as \cite{Hu_tomo}
\begin{equation}
P_{ij}^{\kappa}(\ell) = 
\int_0^{\infty} dz 
\underbrace{\mystrut{3.0ex}\,{W_i(z)\,W_j(z) \over r(z)^2\,H(z)}}_{\rm geometry}\,
\underbrace{\mystrut{3.0ex} P\! \left ({\ell\over r(z)}, z\right )}_{\rm growth},
\label{eq:P_kappa}
\end{equation} 
where each integer multipole $\ell$ corresponds to angular scale of about
$180^\circ/\ell$. Here $r(z)$ is the comoving angular diameter distance,
$H(z)$ is the Hubble parameter, the weights $W_i$ are given by $W_i(\chi(z)) =
{3\over 2}\,\Omega_M\, H_0^2\,g_i(\chi)\, (1+z)$ where $g_i(\chi) =
r(\chi)\int_{\chi}^{\infty} d\chi_s n_i(\chi_s) r(\chi_s-\chi)/r(\chi_s)$, and
$n_i$ is the comoving density of galaxies if the coordinate distance to source
galaxies $\chi_s\equiv\chi(z_s)$ falls in the distance
range bounded by the $i$th redshift bin and zero otherwise. Therefore, weak
lensing probes the growth directly via the redshift-dependence\footnote{The
  separation of the growth and geometry dependencies outlined in
  Eq.~(\ref{eq:P_kappa}) is slightly inaccurate, since the number density of
  source galaxies, $n_i(\chi_s)$, which is in the weights $W_i(z)$,
  technically falls in the growth category.}  of the power spectrum $P(k, z)$
in Eq.~(\ref{eq:P_kappa}).

The statistical uncertainty in measuring the shear power spectrum is
\begin{equation}
\Delta P_{ij}^\kappa(\ell) = \sqrt{\frac{2}{(2\ell+1)f_{\rm sky}} } \left[
  P_{ij}^\kappa (\ell) + \delta_{ij}{\langle \gamma_{\rm int}^2\rangle \over \bar{n}_i}
  \right]~~,
\label{eqn:power_error}
\end{equation}
where $f_{\rm sky}$ is the fraction of sky area covered by the survey and
$\delta_{ij}$ is the Kronecker delta function. The
first term in brackets, which dominates on large scales, comes from cosmic
variance of the mass distribution, and the second, shot-noise term results
 both  from the variance in galaxy ellipticities (``shape noise'') and from
shape-measurement errors due to noise in the images. Therefore, to achieve the 
best weak lensing measurements, we aim to maximize sky coverage
(i.e.\ maximize $f_{\rm sky}$); to minimize the shape noise $\langle
\gamma_{\rm int}^2\rangle$; and to be able to  theoretically model and
experimentally measure shear to as small a scale (high $\ell$) as possible.

\begin{figure}[!t]
\begin{center}
\includegraphics[width=0.60\textwidth]{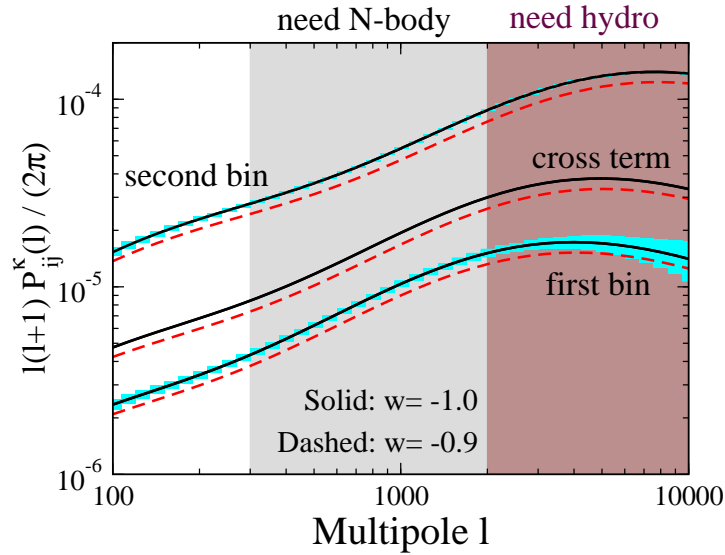}
\end{center}
\caption{ Cosmic shear angular power spectrum and statistical errors expected
  for a future survey such as the LSST. We show two dark energy models with
  equations of state $w=-1$ and $-0.9$; modified gravity theories will also be
  probed but are not shown. For illustration, results are shown for source
  galaxies in two broad redshift bins, $z_s=0$--$1$ (first bin) and
  $z_s=1$--$3$ (second bin); with expected good quality of photometric
  redshifts a much finer slicing of source galaxies in redshift may be
  employed. The cross-power spectrum between the two bins (cross term) is
  shown without the statistical errors. Shaded regions show scales on
  which pure gravity and hydrodynamic simulations, respectively, are necessary to model the
  theory; this is further discussed in Sec.~\ref{sec:sim}.  Adaptation of a
  plot from Ref.~\cite{FriTurHut}. }
\label{fig:P_kappa_tomo}
\end{figure}

Figure \ref{fig:P_kappa_tomo} shows the cosmic shear angular power spectrum
and statistical errors expected for a future survey such as LSST.  We show two
dark energy models with equations of state $w=-1$ and $-0.9$; modified gravity
theories will also be probed but are not shown.  For each cosmology, there are
two curves for the auto-correlations of the shears in two different redshift
bins, and one curve for the cross-correlation of the shears between the two
redshift bins.  Depending on the quality of photometric redshifts, a much
finer slicing of source galaxies in redshift may be employed.  The difference
between the two equations of state is much larger than the statistical errors
expected for LSST (or other planned Stage IV surveys). The shaded regions in
Figure \ref{fig:P_kappa_tomo} show regimes in which dark-matter-only
simulations, and hydrodynamical simulations with baryons, respectively will be
required in order to calibrate the theoretical angular power spectrum.

\begin{figure}[!t]
\begin{center}
\includegraphics[width=0.60\textwidth]{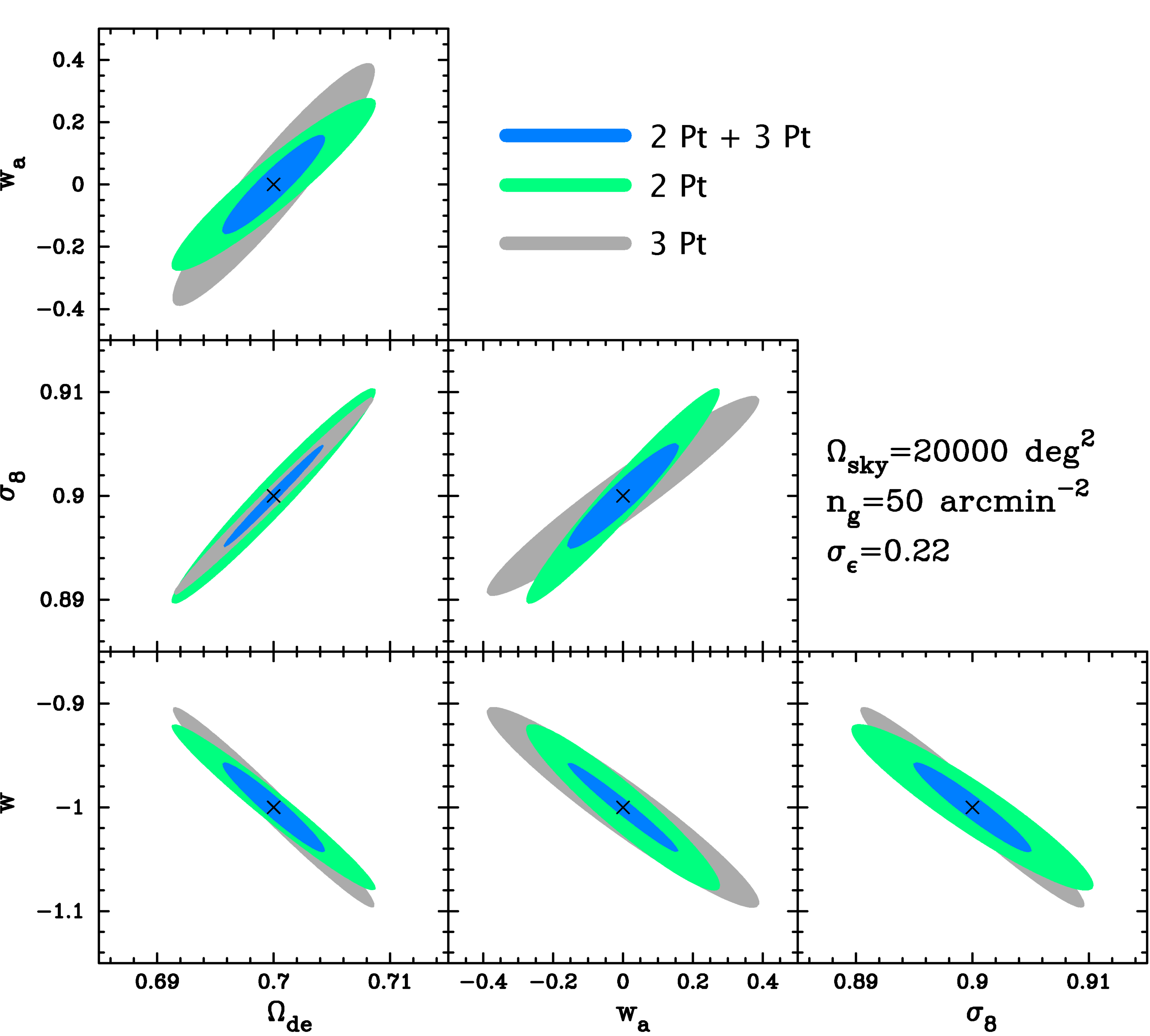}
\end{center}
\caption{Forecasted constraints on cosmological parameters, assuming
  LSST's weak lensing data.  We show the 68\% confidence limit contours for the
  two-point and three-point correlation functions separately (green and gray,
  respectively), as well as the combined constraints using both measurements
  (blue).  These assume priors on the spectral index $n_s$, physical baryon
  density $\Omega_b h^2$ and scaled Hubble constant $h\equiv H_0/(100 {\rm km/s/Mpc})$ 
  from the \Planck\ mission.  Adopted from Ref.~\cite{JarvisTakada05}.  }
\label{fig:two_plus_three}
\end{figure}

In addition to the angular power spectrum, other statistics have been
developed, which have only a somewhat lower statistical power, but which cut
through parameters space differently, so combining them with the power
spectrum can produce significantly better overall constraints on dark energy
parameters.  Figure~\ref{fig:two_plus_three} shows the improved dark energy
constraints that are possible by combining the three-point shear correlation
function (related to the dark matter bispectrum) with the two-point
correlation function (related to the power spectrum).  Because the contours
are somewhat orthogonal to each other in parameter space, the resulting
combined constraints are much better than either one individually.  The
constraints shown in Figure~\ref{fig:two_plus_three} take advantage of the
ability to measure photometric redshifts for the lensed galaxies.  Not
only does this improve the calibration of the source population compared to
what is possible with current surveys, but it also allows us to use multiple
redshift slices (five in this example) for the auto- and cross-correlation
measurements.

The three-point function is merely the simplest statistic that probes the
non-Gaussianity of the underlying dark matter distribution.  Others include
lensing peaks which provide information similar to that provided by galaxy
clusters but are sensitive to all mass, (e.g.\ \cite{Kratochvil:2009wh}),
moments of the convergence distribution, and Minkowski functionals
(e.g.\ \cite{Kratochvil:2011eh}).  These statistics show excellent potential
to improve the weak lensing power spectrum constraints on dark energy.

Finally, weak gravitational lensing is particularly useful probe of modified
gravity. Gravitational lensing observations in general are sensitive to the
sum of the two gravitational potentials $\Phi+\Psi$ (see
Eq.~(\ref{eq:grav_pot})), while particle dynamics probes $\Psi$ alone
(e.g.\ \cite{Uzan_Bernardeau,Jain_Zhang,Bertschinger:2011kk}).  Since
modifications to gravity typically affect the two potentials differently,
combination of weak lensing with other cosmological probes can in principle be
used to differentiate modified gravity from dark energy.

\subsubsection{Systematic Errors and Efforts to Control Them}

While potentially extremely powerful, the underlying shear measurements are
subject to a variety of systematic errors.  There are numerous potential sources of
spurious shear, such as the atmospheric PSF, telescope aberrations and
distortions, charge distribution effects in the CCDs, noise rectification
biases in the shear measurements themselves (since the underlying measurement
is intrinsically nonlinear), to name only the most difficult problems.
Obtaining reliable shear estimates has been an ongoing process within the weak
lensing community.  A series of challenges \cite{STEP1,STEP2,Great08,Great10}
have been testing our ability to measure shear to the required accuracy.  So
far, the state of the art has kept pace with the accuracy required for current
surveys.  However, no one has yet demonstrated a pipeline that can reach that
accuracy required for Stage IV surveys.  A new challenge, dubbed
GREAT3\footnote{http://great3challenge.info/} \cite{GREAT3} tests pipelines at the
accuracy required for these upcoming surveys, and also adds new elements of
realism that had been absent in previous challenges.

The interpretation of weak lensing shear measurements is also complicated by
the photometric redshift biases \cite{MaHuHut,hearinetal10}, calibration of
the predictions in the non-linear regime from N-body simulations
\cite{Huterer_Takada,Rudd_Zentner_Kravtsov,Hearin:2009hz}, and non-Gaussian
errors on small angular scales
\cite{Takada:2008fn,Taylor:2012kz,Dodelson:2013uaa}.  Similar to the effort to
improve shape measurements, there has also been a large effort to handle these
effects to the accuracy that will be required for Stage IV surveys. The
requirements are stringent; for example, future surveys need to calibrate the
mean shear in each of the $\sim$10 redshift bins to about 0.1\% {\it relative}
accuracy in order that dark energy constraints not be significantly degraded
\cite{HTBJ}. However, what really helps weak lensing is the possibility of
``self-calibrating'' the systematic errors --- determining a reasonable set of
the systematic error nuisance parameters from the survey concurrently with the
cosmological parameters without appreciable degradation in accuracy on the
latter. With self-calibration, the survey itself is used to partially
calibrate the systematic effects.

Intrinsic alignments of galaxy shapes (e.g.\ \cite{2010AA...523A...1J}) are
a systematic unique to weak lensing, and many methods that are currently
considered promising for tackling it involve marginalizing over parametrized
models for this effect \cite{Joachimi:2009ez,Kirk:2010zk} which, in turn,
degrades how well dark energy and modified gravity parameters can be
constrained \cite{Laszlo:2011sv,Kirk:2011sw}.  Thus, accurate removal of
intrinsic alignments with minimal loss of cosmological information requires
fairly tight priors on the scaling of intrinsic alignments with galaxy
separation, type, redshift, luminosity, and potentially other parameters.
Existing observational constraints
(e.g.\ \cite{2007MNRAS.381.1197H,2011AA...527A..26J}) are limited by the
requirement that the galaxy sample have both (a) shear estimates and (b)
excellent redshift information (either spectroscopic, spectro-photometric, or
very high-quality photo-$z$) for a reasonably high-density galaxy sample over
a contiguous area large enough to find galaxy pairs with separations of $\sim
100h^{-1}$Mpc.  Since several upcoming surveys will provide shape
measurements, additional overlapping datasets with redshifts will be very
useful in providing constraints on intrinsic alignment models that, in turn,
will be used to remove intrinsic alignments from weak lensing measurements by
Stage IV surveys.  For example, a significant step forward from our current
knowledge would come from a survey with DEEP2-like parameters ($\sim 3$
galaxies$/$arcmin$^2$ for a fairly-selected sample at $z>0.7$) but with ten
times the area.  Since such a survey would be extremely expensive to carry out
spectroscopically, spectro-photometric or many-band photometric surveys might
be the best option for collecting such a dataset.  Note that the requirements
are similar to those for spectroscopic samples for photo-$z$ calibration,
because we need to span the range of different galaxy types; we do however
need a fairly large, contiguous field.

\subsection{Weak Lensing of the Cosmic Microwave Background}

Just as weak gravitational lensing by foreground structure distorts the shape
of background galaxies, it also distorts the fluctuations in the cosmic
microwave background.  In CMB temperature maps, this has the effect of
slightly deflecting trajectories of the CMB photons, which in turn  distorts
the shapes and sizes of the cold and hot spots. In the CMB power spectrum,
lensing smoothes the peaks and also induces
non-gaussian mode coupling that can be measured via the four-point correlation
function.  One can readily construct estimators that can be applied to CMB
temperature maps to reconstruct the map of the deflection field, which in turn
determines the matter distribution and the matter power spectrum integrated over the
observed line-of-sight.  For a review, see \cite{Hanson:2009kr}.

The CMB lensing power spectrum was measured via the four-point function signal
by Atacama Cosmology Telescope (ACT; \cite{Das:2010ga,Das:2013zf}), South Pole
Telescope (SPT; \cite{vanEngelen:2012va}), and most recently by Planck
\cite{Ade:2013tyw}.  The Planck lensing measurement has a lensing detection
significance of 25$\sigma$ and a $5\%$ constraint on the matter power
spectrum.  This translates to a $2.5\%$ constraint on $\sigma_8$ in this
higher redshift range.  Future CMB temperature measurements by ACTpol and
SPTpol, and their upgrades, will improve these constraints by a factor of two
\cite{Niemack:2010wz}.

The angular power spectrum of the {\it deflection} field obeys equation very
similar to Eq.~(\ref{eq:P_kappa}), where contributions along the line of sight
are a product of a geometrical term and the matter distribution, the latter of
which encodes the growth of cosmic structure.  The CMB lensing signal is most
sensitive to matter in the redshift range $z\sim 2$-$4$, which is where its
window functions, corresponding to $W(z)$ in Eq.~(\ref{eq:P_kappa}), peaks.
Thus CMB lensing provides an important anchor at high redshifts for growth
measurements and tightens dark energy and neutrino mass constraints from
lower-redshift growth probes. 

For example, in the Doran-Robbers early dark energy model \cite{Doran:2006kp},
the early dark energy density $\Omega_e$ (defined as dark energy density
relative to critical at $z\gg 10$) can be mimicked by standard dark energy
with a time varying equation of state at late times; the two contributions are
indistinguishable by late-time experiments. CMB lensing measurements can break
this degeneracy: the combined Planck and ground-based Stage III CMB lensing
experiment's data constrain $\Omega_e$ to about one quarter of a percent
($\sigma(\Omega_e) = 0.0025$), while simultaneously constraining the sum of
neutrino masses to 90 meV (rather than 165 meV from Planck alone in this
model) \cite{Calabrese:2010uf}. Note that since both early dark energy and
neutrino mass suppress early growth, constraints on neutrino mass tend to be
tighter in non-early dark energy models; thus early dark energy models give
more conservative bounds.

In addition to generating the lensing signal in CMB temperature maps, lensing also
distorts the polarization of the microwave background, turning E-mode
polarization into B-mode polarization.  This B-mode lensing signal was
recently detected by SPTpol \cite{Hanson:2013hsb}.  While achieving a lensing
detection in polarization maps requires better instrument sensitivity than is
needed for temperature maps, the signal is cleaner since there are fewer
polarized foregrounds.  Sub-percent level constraints on $\sigma_8$ from the
B-mode lensing signal should be within reach of Stage III ground-based CMB
experiments.


\section{Simulations}
\label{sec:sim}
\subsection{Cosmological Simulations of Growth of Structure}

Simulating the dynamical evolution of a representative volume of the
observable Universe, using either particles or grids to model relevant fields,
is an essential method for understanding the non-linear growth of structure
(see \cite{Kuhlen12} for a recent review).  Cosmological simulations enable
the generation of synthetic sky catalogs of galaxies or galaxy clusters with
different levels of observational complexity.  Such synthetic data is now
regularly used to interpret survey results, especially to better understand
issues related to sample/cosmic variance, projection effects, error
covariance, other sources of statistical and systematic uncertainty.  In this
subsection, we discuss the two main modes of dynamical simulations --- N-body
with only gravity and hydrodynamical with baryonic physics --- and comment on
their utility to survey programs.

\subsubsection{Gravity-only N-body Simulations}

N-body simulations evolve the gravitational dynamics of clustered matter,
under an implicit assumption that baryons exactly trace the dark matter on the
resolved scales of the simulations.  This simplifying assumption has the
advantage that the calculation is fast and scales efficiently on parallel
platforms; to date, simulations with nearly a trillion particles in a volume
of several cubic giga-parsec have been conducted, e.g.\ DEUS FUR
\cite{Alimi12}, Horizon Run 3 \cite{Kim11}, and Millennium-XXL \cite{Angulo12}
and the trillion-particle milestone was reached in late 2012 \cite{Habib2012,
  Ishiyama2012}.

N-body simulations are essential for modeling the structure growth at 
trans-linear and non-linear scales where linear perturbation theory
breaks down and higher-order perturbation theory is difficult to perform.  In
fact, a comparison between N-body simulations and higher-order perturbation
theory can be used to cross-check the validity of both methods
\cite{Carlson09}. One of the most important predictions of N-body simulations
is how the matter power spectrum depends on cosmological parameters
\cite{Heitmann10}; as stated in Sec.~\ref{sec:WL}, an accurate prediction of
the matter power spectrum is essential for galaxy clustering and weak lensing
shear correlations. In addition, the halo mass function, which was described
in Sec.~\ref{sec:clusters}, relies on N-body simulations for precision
calibration \cite{ShethTormen99,Jenkins01,Reed03,Lukic07,Tinker08,Crocce10}.
It has been shown that for Stage III dark energy experiments, percent-level
accuracy in mass function is required to avoid severe degradation of dark
energy constraints \cite{CunhaEvrard10,Wu10MF}; achieving this level of
accuracy will require improved simulations of baryon evolution.

Beyond the standard $\Lambda$CDM model, N-body simulations of non-standard
extensions include explorations of quintessence models
\cite{Alimi09,Bhattacharya11}, modified gravity
\cite{Oyaizu_2009,Schmidt09,Chan_Scocc_2009,Khoury_Wyman,Li_Zhao_Tessyer_Koyama},
coupled dark energy and dark matter \cite{Baldi_coupled,Cui12b}, and
self-interacting dark matter \cite{KodaShapiro11}. These simulations provide
us with insights of how these different models affect the growth of structure,
which can be imprinted in the halo mass function, halo bias, and the density
profile of halos.  A challenge in this area, at least on small scales, is that
effects from subtle modifications to the expansion history and linear growth
rate may be degenerate with modifications to baryon physics behavior.

Galaxies and galaxy clusters form in high-density, virialized regions defined
by the dark matter halo population.  To link the outputs of N-body simulations
to observable quantities, a common approach is to derive empirical scaling
relations between halo properties (typically mass or circular velocity) and an
observable property of the halo (e.g.\ central galaxy luminosity or stellar
mass, or cluster X-ray luminosity).  An example of such an approach, the
method of sub-halo abundance matching, has been shown to successfully
reproduce the low-order clustering properties of galaxies observed over a wide
range of redshifts \cite{Conroy06, Reddick13}.  Alternatively, one can trace a
halo's growth history in a simulation and use that behavior, coupled with
rules for internal baryon processing, to predict baryon properties over time.
This so-called semi-analytic approach has seen good success, but increasingly
complex models with large sets of control parameters are necessary to match a
wide range of observations \cite{Baugh06, Benson10}.  On the larger mass
scales of galaxy clusters, one can also apply models for the hot gas
distribution to predict observable X-ray and SZ signals \cite{Angulo12}.
 
N-body codes are largely mature.  Comparisons of independent codes demonstrate
percent-level agreement on the large-scale matter power spectrum
\cite{Heitmann10} and the virial scaling (relation between velocity dispersion
and mass) of dark matter halos \cite{Evrard08}.  Despite their mature status,
N-body simulations will always be limited in their applicability to reality by
the fact that baryons do not trace dark matter on strongly non-linear scales
which, recall, {\it roughly} corresponds to scales of less than a few Mpc.
Inclusion of realistic baryonic physics in cosmological simulations lies at 
the frontier of computational cosmology.

\subsubsection{Hydrodynamical Simulations}

As mentioned in Sec.~\ref{sec:WL}, there is rich dark energy information at
small spatial scales, but one needs an accurate model for small-scale
structure evolution in order to mine this territory productively.  To model
the the growth of small-scale structure, it is essential to understand a
multitude of baryonic processes, including radiative cooling, star and compact
object formation, and feedback from supernovae and active galactic nuclei
(AGN); see \cite{BorganiKravtsov12} for a recent review.  This class of
simulation is, in general, much more computationally intensive than N-body
simulations, and the modeling of star formation and feedback processes is not
yet well understood.  For example, in high mass halos the central mass density
profile can become more concentrated due to star formation or less
concentrated by AGN feedback
\cite{Gnedin04,Rasia04,Tissera10,Martizzi12,Newman13}.  While initial studies
of how baryonic processes alter the matter power spectrum
\cite{Jing06,Rudd08,vanDaalen11,Casarini12} and the halo mass function
\cite{Rudd08,Stanek09,Cui12} have been done, more work is needed to
meaningfully constrain the small scale matter power spectrum and its evolution
over cosmic time.

On very small scales, there have been several long-standing discrepancies
between the structure predicted by $\Lambda$CDM and observational evidence,
including the inner-slope of low-mass galaxy halos (cusp vs.\ core problem)
and the number of satellite galaxies (missing satellite problem); see
\cite{Weinberg13} for a recent review.  It has recently been shown that the
gravitational back-reaction on dark matter driven by small-scale baryonic
feedback can solve these apparent discrepancies \cite{Governato12,Teyssier13}.
The largest halos are relatively immune to galaxy feedback, and early
hydrodynamic simulations of purely gravitational evolution produced X-ray and
SZ properties of the hot gas in galaxy clusters in reasonable agreement with
observations \cite{Evrard90}.  Modern simulations are struggling to reproduce
the low observed fraction of baryons that form stars, but AGN feedback
mechanisms appear promising \cite{BorganiKravtsov12}.

Hydrodynamical simulations produce smaller halo samples than dark matter
simulations, so results are often limited by sample variance.  Because of
astrophysics uncertainties, relatively little attention has been paid to
hydrodynamical simulations in modified gravity models or alternate dark energy
models.

\subsubsection{Synthetic Skies}

Synthetic galaxy catalogs based on N-body simulations are becoming an
indispensable guide for science analysis of large-angle photometric and
spectroscopic surveys \cite{Evrard02,
  Eke04,Yan04,Balaguera12,Font-Ribera12,Sousbie08,Gerke12,Merson13,Cai09}.
Such catalogs provide truth tables that can be processed through selection
machinery to generate survey-specific expectations.  This process can help
guide survey strategy and plan follow-up campaigns, as well as enable insights
into sources of systematic errors in science analysis.  Recent examples
include covariance estimates for SDSS-III BOSS clustering analysis based on
the synthetic catalogs of \cite{Manera13}, and the support of galaxy group
analysis in the WiggleZ spectroscopic survey from the GiggleZ simulations
\cite{Parkinson12}

The increasing sensitivity and sky coverage of galaxy surveys will only
increase the demand for high-fidelity, multi-wavelength synthetic sky maps.  In
particular, catalog-level expectations derived from a simple observational
transfer function may be insufficient.  The ultimate approach would
incorporate the propagation and acquisition of source photons, and their
subsequent conversion to detector signals.

\subsection{Simulating a New Generation of Cosmological Probes}
\label{sec:rationale}

The stringent requirements described above for the control of
systematic errors and uncertainties within cosmological experiments
necessitate a detailed understanding of the properties of individual
experiments or surveys.  Systematic effects can arise from the design
of the system (e.g.\ ghosting of images or scatter light), from the
response of the atmosphere (e.g.\ the stability of the
point-spread-function or the variability in the transmissivity of the
sky), from the strategy used to survey the sky (e.g.\ inhomogeneous
sampling of astronomical light curves), or from limitations in an
analysis algorithms (e.g.\ due to the finite processing power
available for characterizing the properties of detected
sources). Understanding which of these issues will impact the science
(and how) is critical if we hope to maximize our scientific returns.

Over the last few years, simulation frameworks have demonstrated that they can
provide such a capability; delivering a virtual prototype against which design
decisions, optimizations (including descoping), and trade studies can be
evaluated \cite{Connolly12}. What defines the range of capabilities and
fidelity required for a simulation framework? There are clearly trade-offs
between engineering tools, end-to-end simulators, and the use of extant data
sets.  Engineering simulations such as Zemax are typically used to define the
optical design of the system. While detailed, these modeling tools do not
couple to the astrophysical properties of the sky nor the variations in
observing conditions. They are not designed to scale to the size of large
scale experiments or surveys. Extant data sets, in contrast, provide a
representative view of the complexity of observations and the Universe as a
whole. They are, however, constrained by the fact that they represent an
existing experiments and any inherent systematics might not reflect the design
of a new experiment.  For example, in the case of the LSST \cite{lsst}, the
science requirements specify levels of accuracy in characterizing the
photometric, astrometric and shape properties of stars and galaxies that are
between a factor of two and one hundred times better than current surveys
\cite{ivezic}.

\subsubsection{Design Through Simulation}

Instrument simulators, coupled to models of the observable Universe and to
simulations of the cadence of a survey (i.e.\ the time and positional
dependence of a sequence of observations) can provide data with the expected
characteristics of a survey well in advance of first light. Detailed
simulations of the design of a telescope, its optics, or the performance of a
camera or spectrograph can identify the need for new calibration and software
development efforts early in the process, thus enabling a project to prioritize
the development effort to match the science requirements drivers (and identify
which science aspects of the survey were insensitive to these effects). A
simulation framework provides the ability to take a high level requirement,
which incorporates optical-mechanical, atmospheric, electronic, and software
components together with the underlying astrophysical distributions of
sources, and evaluate which systematics are most sensitive to individual
components (i.e. assuming we can model the simulation components at the
appropriate level of fidelity). A simulation framework can provide an
end-to-end implementation of the full flow of photons and information to
evaluate the ability to achieve the science requirements or a simplification
of the flow of information to identify the sub-components and their
contribution to the overall performance.

There are a number of historical instances whereby the design of a survey
(including the analysis software) has impacted the ability of that system to
achieve, in a timely fashion, its stated photometric and astrometric
performance. For example, for the case of the SDSS, the photometric
performance of this system achieved better than 2\% photometric calibration
across its survey volume. To achieve this level of fidelity required the
identification and correction of a number of features impacting the
photometric performance. It was recognized, three years after first light,
that an accurate model for the point-spread-function of the SDSS telescope and
its variation across the focal plane needed to be developed \cite{ivezic03}.
After five years, techniques for a global photometric solution for the SDSS
photometry were implemented in order to obtain a 2\% photometric calibration
\cite{Padmanabhan}. Simulations provide the capability to address many of
these issues prior to operations.

\subsubsection{Performance Verification}

During the preconstruction and construction phases of any experiment,
prototype devices and subcomponents will be delivered together with laboratory
data on the performance of these systems (e.g.\ the delivery of sensors with
measured quantum efficiencies, defects, and noise characteristics). Evaluating
the impact of these components, prior to the completion of construction, is a
non-trivial task.  Engineering models and simulations provide some of these
capabilities (e.g.\ the use of FRED to evaluate integrated scattered light).
Laboratory measurements do not, however, equate directly to the science
capabilities as we must couple the performance of a device with the properties
of astrophysical sources and our ability to measure the properties of sources
to characterize and correct for any systematic effects.

\subsubsection{Diagnostics and Trade Studies}

Science requirements propagate into scientific analyses. For example, the
photometric redshifts, which are relied upon by many cosmological probes, depend
on deblending of sources, photometric zero points, and the implementation of
model-based magnitudes. Typically the requirement on the photometric redshift
performance captures only the final level of fidelity (e.g.\ the variance or
fraction of outliers in the redshift relation). Simulations, where the input
configurations can be controlled, in conjunction with observational data sets,
provide the ability to quantify the sensitivity of these requirements to the
input parametrization of the Universe and the properties of the site. Trade
studies, such as the impact of available compute resources on the fidelity of
the derived shape parameters, can be undertaken in controlled situations to
define what governs the sensitivity of the system.

\bibliography{growth}


\end{document}